\documentclass[sigconf]{acmart}

\usepackage{listings}
\usepackage{multirow}
\usepackage{pifont}
\usepackage{threeparttable}
\usepackage{pbox}
\usepackage{caption}
\usepackage{color, colortbl}
\usepackage{xspace}
\usepackage{amsmath}
\usepackage{caption}
\usepackage{subcaption}
\usepackage{algorithm}
\usepackage{algpseudocode}
\usepackage{enumitem} 
\usepackage{adjustbox}
\usepackage{tikz}
\usepackage{soul}
\usepackage{booktabs}
\usepackage[titletoc]{appendix}

\newcommand{\ptrix}{{\sc PTrix}\xspace}

\newcommand{\etal}{\mbox{\emph{et al.\ }}}
\newcommand{\ie}{\mbox{\emph{i. e.,\ }}}
\newcommand{\eg}{\mbox{\emph{e.g.,\ }}}
\newcommand{\afl}{{\tt AFL}\xspace}

\newcommand{\point}[1]{\par\vspace{0.1in}\noindent{\bf #1:}}
\newcommand{\rom}[1]{\textit{\expandafter\romannumeral #1}}
\newcommand\todo[1]{\color{red}TODO:~#1}

\newcommand{\numbug}{{\textbf{35}}\xspace}
\newcommand{\nummembug}{{\textbf{25}}\xspace}
\newcommand{\numdosbug}{{\textbf{10}}\xspace}
\newcommand{\numcve}{{\textbf{11}}\xspace}

\algnewcommand\algorithmicinput{\textbf{INPUT:}}
\algnewcommand\algorithmicoutput{\textbf{OUTPUT:}}
\algnewcommand\INPUT{\item[\algorithmicinput]}
\algnewcommand\OUTPUT{\item[\algorithmicoutput]}

\floatstyle{ruled}
\newfloat{code fragment}{thp}{lop}
\floatname{code fragment}{Code Fragment}

\definecolor{dkgreen}{rgb}{0,0.6,0}
\definecolor{gray}{rgb}{0.5,0.5,0.5}
\definecolor{mauve}{rgb}{0.58,0,0.82}
\def\BibTeX{{\rm B\kern-.05em{\sc i\kern-.025em b}\kern-.08emT\kern-.1667em\lower.7ex\hbox{E}\kern-.125emX}}
    
\copyrightyear{2019}
\acmYear{2019}
\setcopyright{acmlicensed}
\acmConference[AsiaCCS '19]{ACM Asia Conference on Computer and Communications Security}{July 9--12, 2019}{Auckland, New Zealand} 
\acmBooktitle{ACM Asia Conference on Computer and Communications Security (AsiaCCS '19), July 9--12, 2019, Auckland, New Zealand}
\acmPrice{15.00} 
\acmDOI{10.1145/3321705.3329828}
\acmISBN{978-1-4503-6752-3/19/07}

\begin{document}

%
% The "title" command has an optional parameter, allowing the author to define a "short title" to be used in page headers.
\title{\ptrix: Efficient Hardware-Assisted Fuzzing for COTS Binary}

% \author{}

%
% The "author" command and its associated commands are used to define the authors and their affiliations.
% Of note is the shared affiliation of the first two authors, and the "authornote" and "authornotemark" commands
% used to denote shared contribution to the research.

%\renewcommand*{\thefootnote}{\fnsymbol{footnote}}

\author{Yaohui Chen*}
\thanks{*These two authors have contributed equally}
\affiliation{%
 \institution{Northeastern University}
%  \streetaddress{P.O. Box 1212}
}

\author{Dongliang Mu*}
\affiliation{%
 \institution{Penn State University}
%  \streetaddress{1 Th{\o}rv{\"a}ld Circle}
%  \country{Iceland}
}

\author{Jun Xu}
\affiliation{%
 \institution{Stevens Institute of Technology}
}

\author{Zhichuang Sun}
\affiliation{%
\institution{Northeastern University}
% \country{India}
}

\author{Wenbo Shen}
\affiliation{%
 \institution{Zhejiang University}
%  \streetaddress{30 Shuangqing Rd}
%  \country{China}
 }

\author{Xinyu Xing}
\affiliation{%
 \institution{Penn State University}
%  \streetaddress{8600 Datapoint Drive}
%  \postcode{78229}
 }

\author{Long Lu}
\affiliation{\institution{Northeastern University}}

\author{Bing Mao}
\affiliation{\institution{Nanjing University}}

%
% By default, the full list of authors will be used in the page headers. Often, this list is too long, and will overlap
% other information printed in the page headers. This command allows the author to define a more concise list
% of authors' names for this purpose.
\renewcommand{\shortauthors}{Chen, et al.}

%  Show the XML Only

%\begin{CCSXML}
%<ccs2012>
%  <concept_id>10010520.10010553.10010562</concept_id>
%  <concept_desc>Computer systems organization~Embedded systems</concept_desc>
%  <concept_significance>500</concept_significance>
%  <concept_id>10010520.10010575.10010755</concept_id>
%  <concept_desc>Computer systems organization~Redundancy</concept_desc>
%  <concept_significance>300</concept_significance>
%  <concept_id>10010520.10010553.10010554</concept_id>
%  <concept_desc>Computer systems organization~Robotics</concept_desc>
%  <concept_significance>100</concept_significance>
%  <concept_id>10003033.10003083.10003095</concept_id>
%  <concept_desc>Networks~Network reliability</concept_desc>
%  <concept_significance>100</concept_significance>
%</ccs2012>
%\end{CCSXML}
%
%\ccsdesc[500]{Computer systems organization~Embedded systems}

%
% Keywords. The author(s) should pick words that accurately describe the work being
% presented. Separate the keywords with commas.

\begin{abstract}
%We show path sensitive can bring different code coverage than edge based fuzzer in short term.
%We also show that using hardware to accelerate a path-sensitive fuzzer can lead to superior performance (i.e, the capability of achieving higher code coverage and finding more bugs). 
%We conduct rigorous experiments to show when using path-sensitive feedback with speed optmization, a much better performant fuzzer is possible.

Despite its effectiveness in uncovering software defects, American Fuzzy Lop
(\afl), one of the best grey-box fuzzers, is inefficient when
fuzz-testing source-unavailable programs. \afl's binary-only fuzzing mode,
 {\tt QEMU-AFL}, is typically 2-5$\times$ slower than its source-available
fuzzing mode. The slowdown is largely caused by the heavy dynamic instrumentation.

Recent fuzzing techniques use Intel Processor Tracing (PT), a light-weight
tracing feature supported by recent Intel CPUs, to remove the need of dynamic
instrumentation. However, we found that these PT-based fuzzing techniques are
even slower than {\tt QEMU-AFL} when fuzzing real-world programs, making them less effective than {\tt QEMU-AFL}. This poor performance is caused by the slow extraction of code coverage information from highly compressed PT traces.

In this work,  we present the design and implementation of \ptrix, which fully
unleashes the benefits of PT for fuzzing via three novel techniques. First, 
\ptrix introduces a scheme to highly parallel the processing of 
PT trace and target program execution. Second, it directly takes decoded PT 
trace as feedback for fuzzing, avoiding the expensive reconstruction of code 
coverage information. Third, \ptrix maintains the new feedback with stronger feedback than 
edge-based code coverage, which helps reach new code space and defects that {\afl} may not. 

We evaluated \ptrix by comparing its performance with the state-of-the-art
fuzzers. Our results show that, given the same amount of time, \ptrix achieves 
a significantly higher fuzzing speed and reaches into code regions missed by
the other fuzzers. In addition, \ptrix identifies \numbug new vulnerabilities 
in a set of previously well-fuzzed binaries, showing its ability to complement existing fuzzers.

\end{abstract}

%
% This command processes the author and affiliation and title information and builds
% the first part of the formatted document.
\maketitle

\section{Introduction}
\label{sec:intro}

%Background

Fuzz-testing, or fuzzing, is an automated software testing technique for
unveiling various kinds of bugs in software. Generally, it provides invalid or
randomized inputs to programs with the goal of discovering unhandled exceptions
and crashes.  This easy-to-use technique has now become the de facto standard in
the software industry for robustness testing and security vulnerability
discovery.

Among all the fuzzing tools, American Fuzzy Lop (\afl) requires essentially no
a-priori knowledge to use and can handle complex, real-world
software~\cite{afl}. Therefore, \afl and its extensions have been widely adopted
in practice, constantly discovering unknown vulnerabilities in popular software
packages (such as {\tt nginx}, {\tt OpenSSL}, and {\tt PHP}).

A major limitation of \afl is its low speed in fuzzing source-unavailable software.
Given a commercial off-the-shelf (COTS) binary, \afl needs to perform a {\em black box on-the-fly} instrumentation using
a customized version of {\tt QEMU} running in the ``user space emulation'' mode.
Despite the optimizations~\cite{bellard2005qemu}, {\tt QEMU}
still incurs substantial overhead in this mode and thus slows down \afl's
binary-only fuzzing. According to the \afl white paper~\cite{afltech}, 
\afl gets decelerated by {\tt 2 - 5$\times$} in this {\tt QEMU}-based mode,
which is significant enough to make \afl much less used for binary-only fuzzing.

% Existing research focus

Previous research primarily focused on improving \afl's code coverage so that it
could potentially find more bugs. To the best of our knowledge, only a few works
aimed to improve the efficiency/speed of \afl~\cite{kafl,designing,honggfuzz}. Since quickly identifying
software flaws can expedite patches and narrow exploit windows of
vulnerabilities, the goal of this work is to improve \afl's efficiency at
uncovering bugs in COTS binaries.

% Our work

Unlike the prior work that achieves efficiency improvement through syscall
re-engineering~\cite{designing}, we propose a new fuzzing mechanism utilizing a
recent hardware tracing feature, namely Intel PT~\cite{intel-pt}, to enhance the performance of binary-only
fuzzing. We design and develop \ptrix, an 
{\em efficient hardware-assisted} fuzzing tool. The intuition of using PT to accelerate fuzzing is as follows. 
The success of \afl is largely attributable to the use of code coverage as feedback. 
To obtain code coverage information, \afl traces the program execution with {\tt QEMU}, 
 which incurs significant overhead. Alternatively, Intel PT can
trace program execution on the fly with negligible overhead. By replacing {\tt
QEMU} with lightweight hardware tracing, we can improve the efficiency for
binary-only fuzzing.

Intel PT stores a program execution trace in the form of compressed  binary
packets. To implement \ptrix, an instinctive reaction~\cite{kafl} is to sequentially
trace the program execution, decode the binary packets, and translate them into
code coverage that \afl needs as feedback. We refer to this implementation as 
{\tt Edge-PT}. However, as we demonstrate in
Section~\ref{sec:eval}, {\tt Edge-PT} introduces significant run-time
overhead to fuzzing and does not actually benefit binary-only fuzzing with
efficiency improvement. This is due to the fact that binary packet decoding and
code translation both incur high computation cost.

To address the above issue, we first introduce a {\em parallel, elastic} scheme to parse a PT trace. 
This scheme mounts a concurrent thread to process the execution trace in parallel with 
the target program execution. Due to a hardware restriction, the boundary of the execution trace
can only be updated when PT is paused. This frequently defers the parsing thread  
until the next boundary update which may arrive after termination of the target program. 
To overcome this limitation, our scheme leverages an elastic approach to automatically adjust the time window 
of target program execution (as well as PT tracing). Our approach ensures that the trace boundary 
gets safely and timely updated and the parsing thread are used efficiently.

Despite the above parallel scheme, we still observe that 
the parsing thread frequently and dramatically falls behind the program execution. 
The major cause is the aforementioned high cost of code coverage reconstruction. 
To this end, we replace the code-coverage feedback used by \afl with a newly 
invented PT-friendly feedback mechanism. Our mechanism directly encodes the stream of
PT packets as feedback. This makes \ptrix
no longer need to perform code coverage reconstruction,
which ultimately enables the parsing thread to accomplish its job almost at the same time as the
target program finishes executing on the fuzzing input. Facilitated by these new designs, 
\ptrix executes {\tt 4.27x} faster than {\tt \afl} running in {\tt QEMU} mode.

Functionality wise, our new feedback does not reduce the guidance that code coverage can provide. 
In essence, the stream of PT packets keeps track of the execution paths, which 
carries not only information about code coverage but also orders and combinations among code block transitions.
This means our new feedback is inclusive of that used by \afl. As we demonstrate in Section~\ref{sec:eval}, 
our feedback allows \ptrix to cover code space quicker, explore code chunks that would otherwise have not been touched, 
and follow through long code paths to unveil deeply hidden bugs.
By the time of writing, \ptrix has identified \numbug previously unknown 
security defects in well-fuzzed programs. 

We note that this work is not the first that applies Intel PT to fuzz
testing~\cite{kafl, honggfuzz, winafl}. To the best of our knowledge,  \ptrix, however, is the first work that
explores Intel PT to accelerate fuzzing. Going beyond the higher efficiency it brings, 
\ptrix also exhibits better fuzzing effectiveness and new
ability to find unknown bugs. While our prototype of \ptrix is built upon
Linux on x86 platform, our design can be generally applied to other operating
systems across various architectures which also support hardware-assisted
execution tracing.

In summary, this paper makes the following contributions.

\begin{itemize}  

	\item We explored Intel PT and utilized it to design an {\em efficient}
	{\em hardware-assisted} fuzzing mechanism to improve efficiency and effectiveness 
	for binary-only fuzzing.

	\item We prototyped our proposed fuzzing mechanism with \ptrix on Linux and
	compared it with other fuzzing techniques, demonstrating it can accelerate a
	binary-only fuzzing task for about {\tt 4.27}$\times$.

	\item We devised a rigorous evaluation scheme and showed: (\rom{1}) Intuitively
	applying PT does not produce an efficient binary-compatible fuzzer;
	(\rom{2}) \ptrix not only improves fuzzing efficiency but also	
	has the potential to explore deeper program behaviors. As of the preparation
	of this paper, \ptrix has identified \numbug unknown
	software bugs, \numcve of them have CVE IDs assigned.

\end{itemize}

% Roadmap
\iffalse
The rest of the paper is organized as follows. In Section~\ref{sec:bg}, we
outline the technical background for understanding the architecture design of
\ptrix and briefly describe how the path-sensitive scheme works. In
Section~\ref{sec:design} and Section~\ref{sec:impl}, we discuss \ptrix's design
and implementation in detail. We evaluate its efficiency and effectiveness in
Section~\ref{sec:eval}. 
%and discuss future directions in Section~\ref{sec:discuss}. 
Related works are presented in
Section~\ref{sec:related} before we conclude the paper in
Section~\ref{sec:conclude}.
\fi

\begin{figure}[t]
\begin{center}
\includegraphics[scale=0.36]{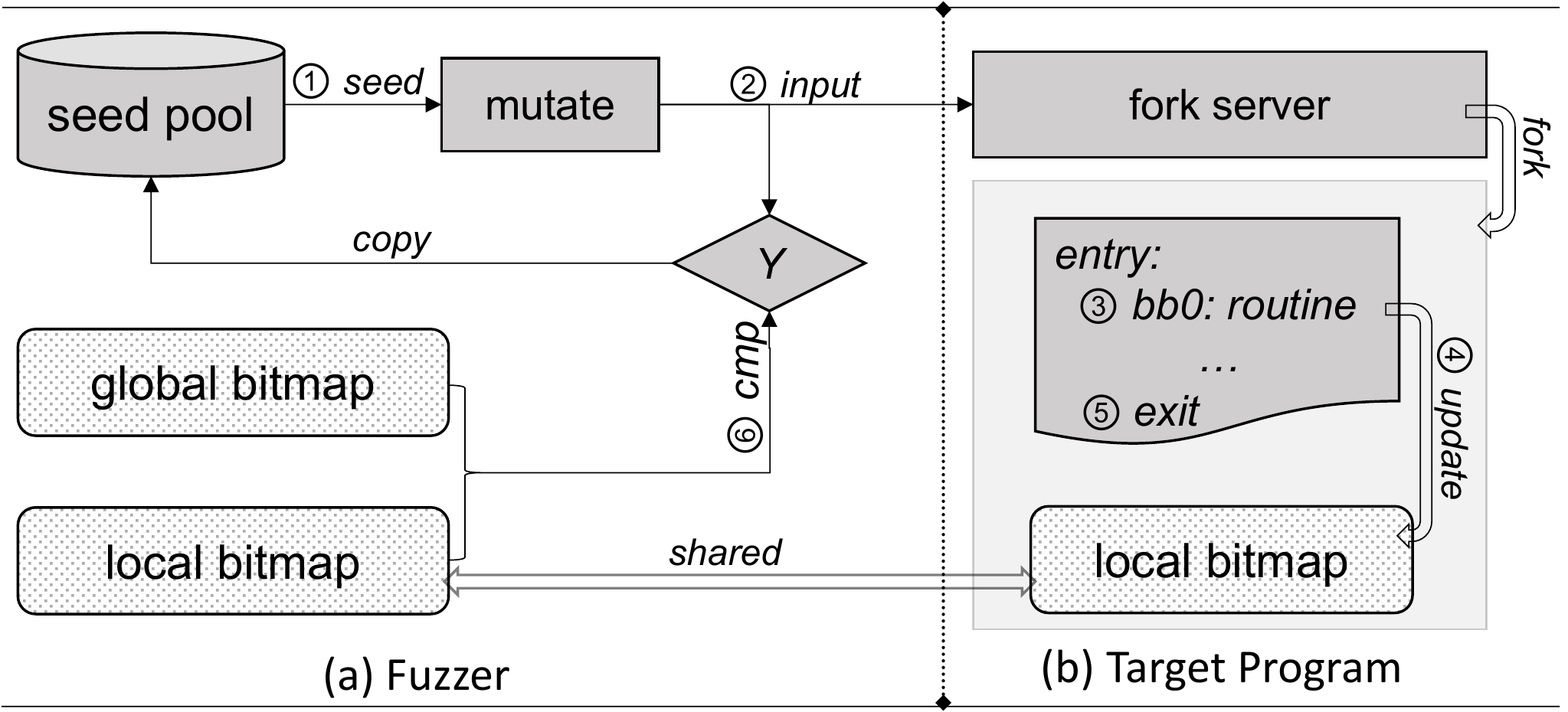}
\caption{The workflow of the fuzzer residing in \afl.}
\label{figs:afl}
\end{center}
\vspace{-1.5em}
\end{figure}

\section{Background}
\label{sec:bg}

Recall that we build \ptrix on top of \afl through Intel PT with the goal of
improving efficiency and effectiveness for fuzzing. In this section, we describe
the background of \afl and that of Intel PT.
 
\subsection{American Fuzzy Lop}

\afl consists of two main components -- an {\em instrumentor} and a {\em
fuzzer}.  Given a target program, the instrumentor performs program
instrumentation by assigning an ID to each basic block (BB) and inserting a
routine at the entry site of that BB. With the routine along with the ID tied to
each BB, the fuzzer follows the workflow below to interact with the target
program and perform continuous fuzz testing.

As is illustrated in Figure~\ref{figs:afl}, the fuzzer starts a fuzzing round by 
scheduling a seed from the pool (\textcircled{1}). It then mutates this seed 
via approaches such as bit-flip to produce new test cases. Using each of these 
test cases as input, the fuzzer launches the target program (\textcircled{2}).  
With the facilitation of the routine instrumented, the target program computes 
hit counts pertaining to the edge indicated by each pair of consecutively executed
BBs (\textcircled{3}) and stores this information to a local bitmap  (\textcircled{4}).
As depicted in Figure~\ref{figs:afl}, the local bitmap is in a memory
region shared by the target program and the fuzzer. 

As is shown in the figure, when the execution of the target
program is terminated (\textcircled{5}), the fuzzer measures the quality of
the input by comparing the information held in the local bitmap with that in the
global one (\textcircled{6}). To be more specific, it examines whether there exists new coverage that
has not yet been observed in the global bitmap. By new coverage, it means the edges 
or the hit counts tied to the edges that have not yet been observed in
previous fuzzing rounds.  For the new coverages identified, the fuzzer
includes them into the global bitmap and then appends the corresponding input to
The fuzzer would then select a new seed for the consecutive rounds of fuzz testing.

To improve the efficiency, as is illustrated in
Figure~\ref{figs:afl}, \afl also introduces a \emph{fork server
mode}~\cite{forksrv}, where the target program goes through {\tt execve()}
syscall and the linking process and then turns to a fork server. 
Then for each round of fuzz testing, the fuzzer clones a new target process 
from the copy-on-write fork server that is perpetually
kept in a virgin state. With this design, \afl could avoid the overhead incurred
by heavy and duplicate execution prefix, and thus significantly expedite the
fuzzing process.

The aforementioned description indicates how \afl works on source-available programs. 
In the situation where source code is unavailable, the aforementioned technique, however, cannot be directly applied to
a target program because binary instrumentation could potentially introduce
unexpected errors. To address this issue, \afl performs dynamic instrumentation
using the user-mode emulator of {\tt QEMU}. Technically, this
design does not vary the fuzzer component residing in \afl. As a result, a
binary-only fuzzing process still follows the workflow depicted in
Figure~\ref{figs:afl}. More details could be referred to at~\cite{afltech}.

% and obtains the coverage information via dynamic instrumentation. 
% The remaining fuzzing logic and workflow are identical to the above stated. 
% As \afl runs the target program via emulation, it reduces the fuzzing efficiency by multiple times. 
% We are thus motivated to design a reliable and efficient fuzzer for binary programs. 

%As \afl runs the target program via, it  
%The above describes how \afl works with source available programs, 
%which, 
% offline static binary rewriting is unreliable and error-prone. 
%{\jun{Will change this paragraph later}} Figure~\ref{figs:afl} illustrates the overview of AFL in fork server mode. In this new scheme, instead of invoking {\tt do\_execve} every time a new input is prepared, the fuzzer sends a \textit{start} message to the fork server process, which then forks a child process to start executing from a check point with the new input. This can work because fork server is compiled as an relocatable (.o) file, which is then linked into the target program. Thanks to this program body sharing design, such checkpoint and resume process can be realized using a simple combination of loop and the {\tt fork} syscall. 
%When the target finish executing, fork server forwards the exit status of the finished child process and then loop back to wait for the start message from fuzzer for the next round of fuzzing.

\begin{figure}[t]
\begin{center}
\includegraphics[scale=0.40]{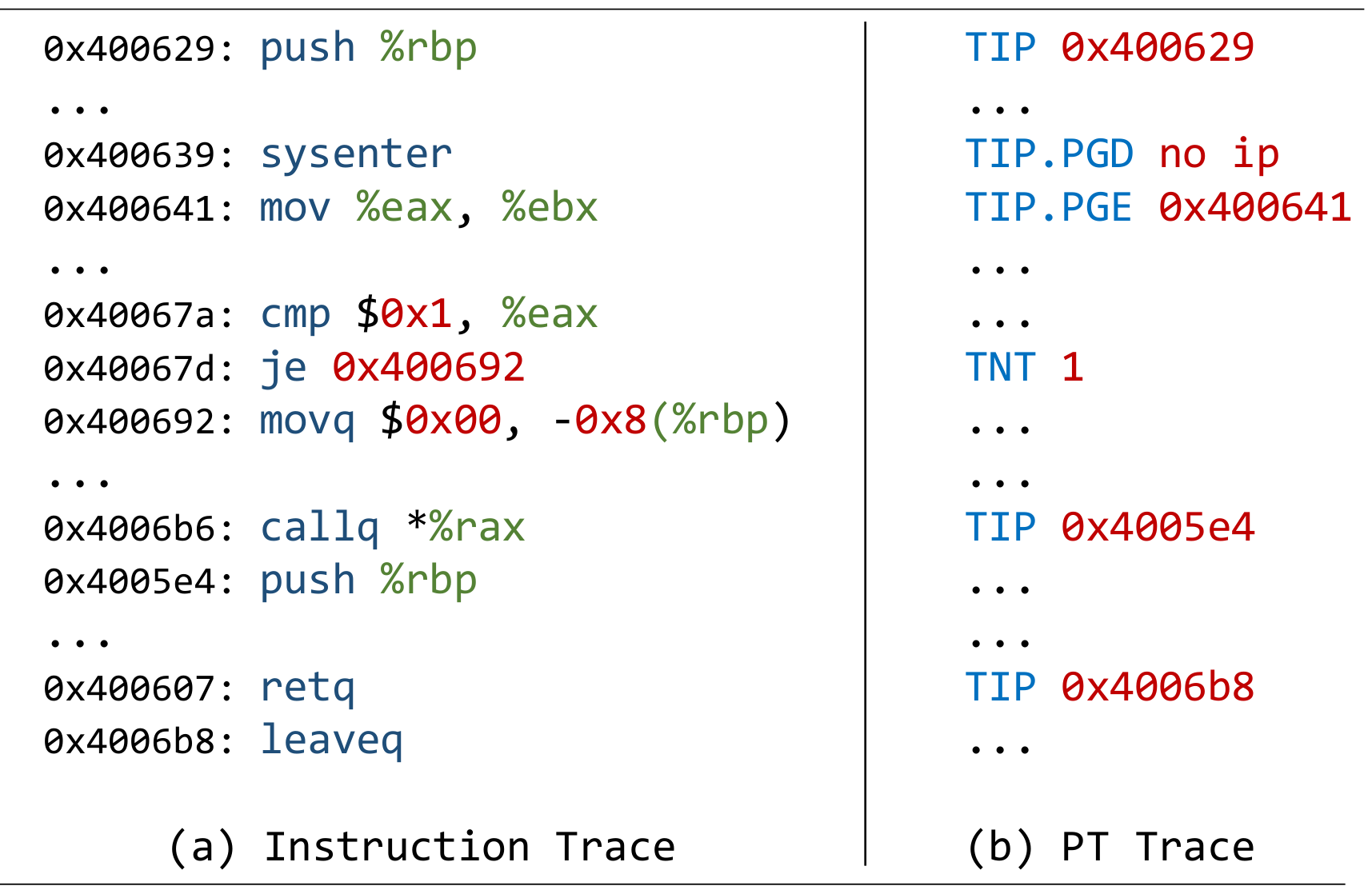}
\caption{Example of trace generated by PT (with kernel tracing disabled). 
The left part shows the instruction sequence and the right part presents the 
corresponding PT trace.}
\label{figs:pt}
\end{center}
\vspace{-2em}
\end{figure}

\subsection{Intel Processor Tracing}

%%%% Trying to add some random stuff so far

Intel PT is a low-overhead hardware feature available in recent Intel processors
(e.g., Skylake series).  It works by capturing information pertaining to
software execution. To minimize the storage cost, Intel PT organizes the
information captured in different forms of data packets. Of all the data
packets, Taken Not-Taken (TNT) and  Target IP (TIP) packets are the ones most
commonly adopted. Technically speaking, TNT packets take the responsibility of
recording the selection of conditional branches, whereas TIP packets are used
for tracking down indirect branches and function returns. Along with some other
packets such as Packet Generation Enable (PGE) and Packet Generation Disable (PGD), Intel PT also utilizes TIP packets to trace
exceptions, interrupts and other events.

%Beside packets for tracing, there are many other types of packets which record the working status 
%of PT itself. For example, Packet Generation Enable (PGE) and Packet Generation Disable (PGD) 
% respectively records where and when PT enables and disables tracing.  

Using the packet trace captured by Intel PT along with the corresponding target
program in the binary form, a software developer or a security analyst could
fully and perfectly reconstruct the instruction trace pertaining to the
execution of the target program. To demonstrate this, we depict the packet trace
as well as the target program in disassembly side by side in
Figure~\ref{figs:pt}. As we can observe from the figure, Intel PT records the
address of the entry point with TIP packet {\tt TIP 0x400629} and then the
conditional jump with a TNT packet indicated by {\tt TNT 1}. Following these two
packets, Intel PT also encloses packets {\tt TIP 0x4005e4} and {\tt TIP 0x4006b8}
in the packet trace. Using the first two packets shown in the trace, we can
easily infer that the program enters its execution at the site {\tt 0x400629}
and then takes the true branch redirecting the execution from the site {\tt
0x40067d} to the site {\tt 0x400692}. As is indicated by consecutive packets
{\tt TIP 0x4005e4} and {\tt TIP 0x4006b8}, we can further conclude that the target
program invokes a subroutine located at the site {\tt 0x4005e4} and then returns
to the site {\tt 0x4006b8}.

% In this case, the true branch is taken 
% (jump from {\tt 0x40067d} to {\tt 0x400692}), and PT correspondingly generates a TNT packet({\tt TNT 1}). 
% Following that, the execution will reach two indirect jumps at {\tt 0x4006b6} and
% {\tt 0x400607}. Respectively, PT records their targets ({\tt TIP 0x05e4} and {\tt TIP 0x0607}). 
% When the control flow is trapped into ({\tt 0x400639}) and return from kernel ({\tt 0x400641}), 
% PT produces two packets ({\tt TIP.PGD} 
% and {\tt TIP.PGE}) which indicate 
% the pause and resume of tracing. 

% The following is saved for later

% For the sake of flexibility, PT provides all-sided configuration options. Of particular interests, 
% PT support filtering based on privilege level and address ranges. 
% This enables \ptrix to selectively monitor the main executable and/or any library of the target program.

%AFL takes edge-coverage based feedback. This type of feedback has certain limitations. 

%guideline:
%novel ideas/designs
%   new parsing: parallelism vs. post-exec parsing
%
%   new coverage representation:
%       more efficient (both time and space) for construct, storage, lookup
%       better differentiakion of exec pathsSection
%
%unique challenges:
%   (design) larger bitmap (path storage):decreasing path sensitivity for hashing
%   (impl.)  
%   (impl.)    
%   (impl.)    
%   (impl.)    

\section{Design}
\label{sec:design}
%YH: could delete
%In this section, we present the system overview of \ptrix. Then, we elaborate
%the workflow of the system, followed by the description about how to improve its
%fuzzing efficiency through a series of systematic and algorithmic designs.
%Finally, we discuss the impact of \ptrix upon discovering unknown
%vulnerabilities and exploring deep code path and subtle program behavior changes.

\begin{figure}[t]
\begin{center}
\includegraphics[scale=0.30,clip]{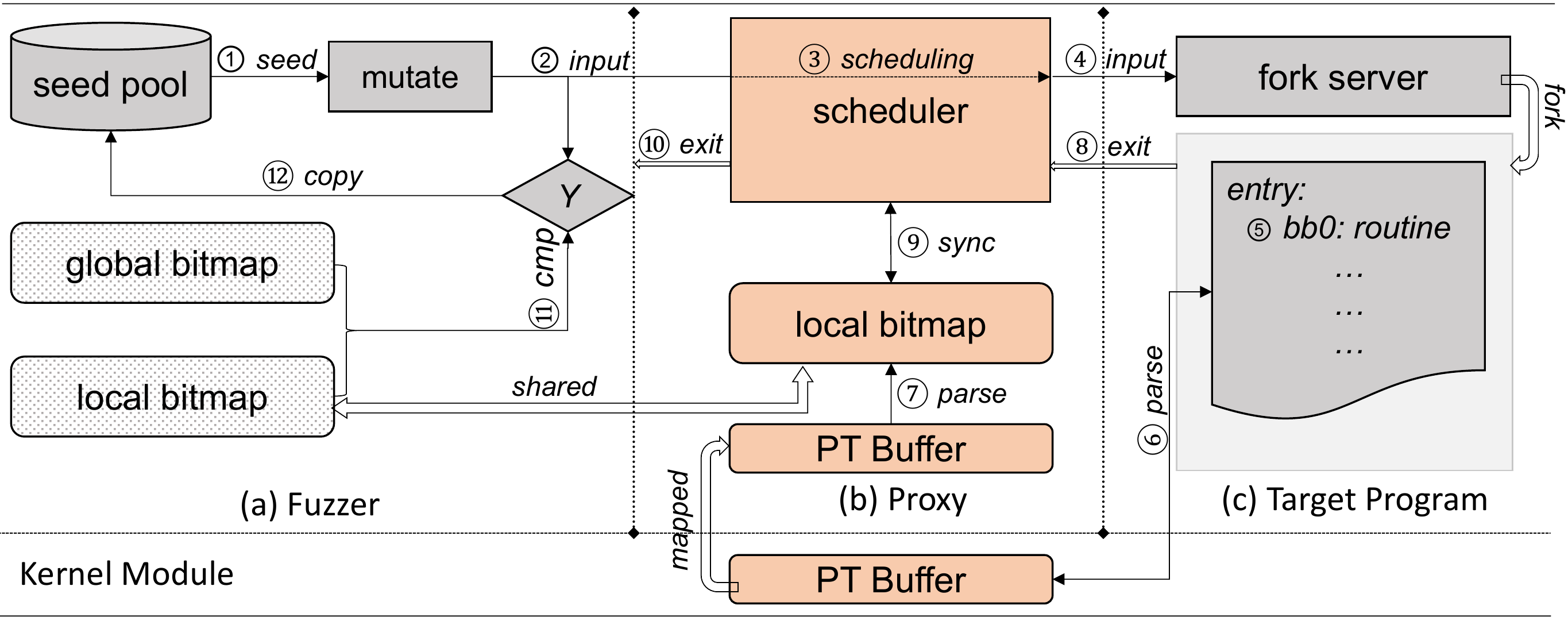}
\caption{System architecture and workflow of \ptrix. The Fuzzer, Proxy, and Target Program 
are separate processes in the user space. The Kernel Module is a driver running inside the kernel space.
Components in orange color are newly introduced by \ptrix.}
\label{figs:overview}
\end{center}
\vspace{-2em}
\end{figure}

\subsection{Overview}
\label{subsec:overview}

As is depicted in Figure~\ref{figs:overview}, \ptrix shares with conventional
\afl the same architecture except for a PT module as well as a proxy sitting 
between the fuzzer and the target program. Within this new fuzzing system, the
proxy component takes the responsibility of coordinating fuzz testing, and the
PT module is used for supporting the parallel and elastic parsing of Intel PT trace packets. 
In the following, we briefly describe how each component coordinates with each other at the high level. Note that a
more detailed description of the workflow will be provided in
Section~\ref{subsec:workflow}.

Similar to \afl, \ptrix starts with generating an input for the
target program (\textcircled{\raisebox{-0.9pt}1}~\textcircled{\raisebox{-0.9pt}2}). Instead of passing the input
directly to the program or more precisely the embedded fork server, \ptrix however sends
it through the proxy component which leverages a scheduler to coordinate fuzz
testing (\textcircled{\raisebox{-0.9pt}3}~\textcircled{\raisebox{-0.9pt}4}).

With the facilitation of Intel PT, \ptrix uses a PT module to monitor the
execution of the target program and store the trace packets in a pre-allocated buffer shared
between kernel and user space (\textcircled{\raisebox{-0.9pt}6}). Carried on simultaneously with the execution of
the target program, the proxy parses the PT trace, computes feedback
and updates the local bitmap accordingly (\textcircled{\raisebox{-0.9pt}7}).

At the time of the termination of fuzz testing, the proxy receives a
notification (\textcircled{\raisebox{-0.9pt}8}). To enforce correct synchronization between
consecutive rounds of fuzz testing, the scheduler of the proxy does not pass the
notification back to the fuzzer until it confirms the completion of packet
parsing (\textcircled{\raisebox{-0.9pt}9}~\textcircled{10}).

On the fuzzer side, right after receiving the fuzzing completion notification from the proxy,
it follows the same procedure as \afl to conclude one round of fuzz testing, i.e.,
comparing the local and global bitmaps and, if necessary, appending the input
into the queue for the consecutive rounds of fuzz testing
(\textcircled{11}~\textcircled{12}). It should be noted that, throughout the
fuzzing process described above, the key characteristic of \ptrix is to compute
path coverage using PT trace. As is mentioned
earlier in Section~\ref{sec:intro}, this could significantly reduce the overhead
introduced by instruction trace reconstruction. In Section~\ref{subsec:efficiency}, we
will elaborate on our design of \ptrix to enable this practice.

\begin{figure}[t]
\hrulefill
\begin{center}
\includegraphics[scale=0.3]{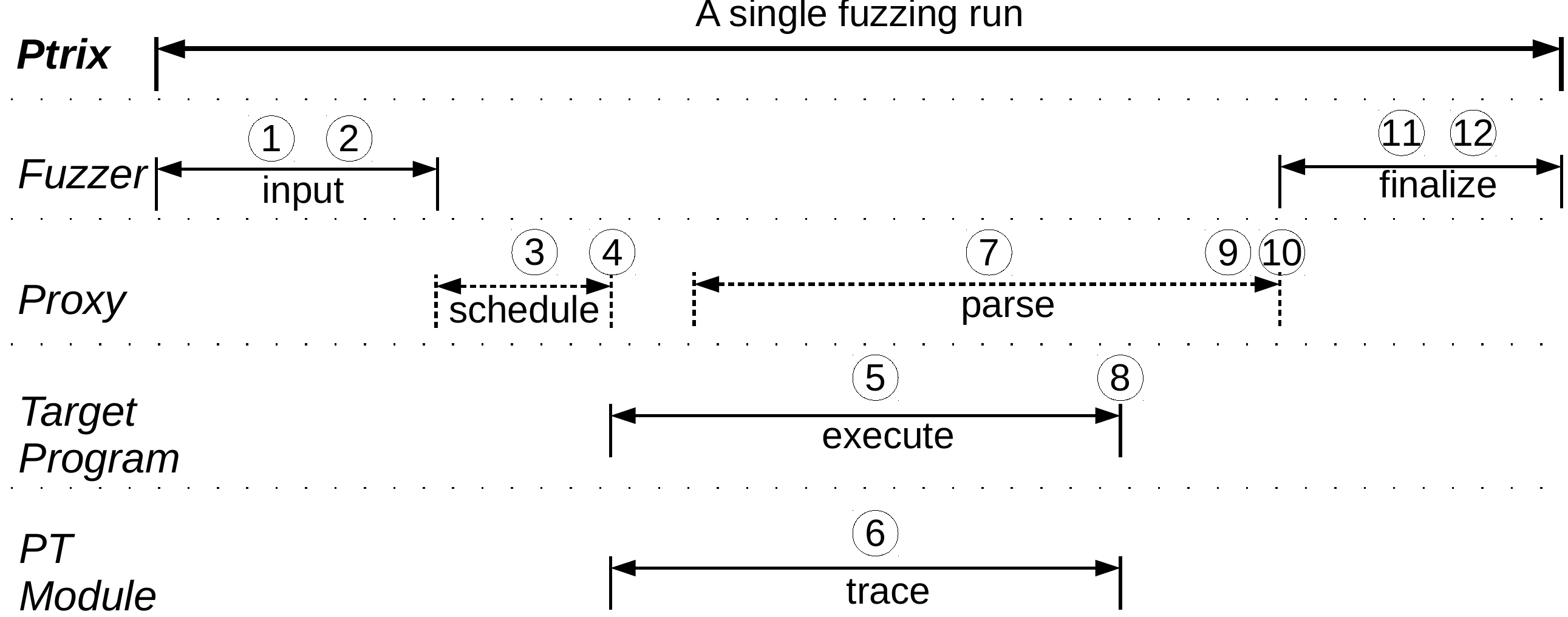}
\end{center}
\vspace{-2.5mm}
\hrulefill
\caption{Timeline of one fuzzing round in \ptrix. Note that the intervals depicted in dotted lines are those we aim to introduce performance optimization, and the circled numbers correspond to those shown in Figure~\ref{figs:overview}.}
\label{figs:timeline}
\end{figure}
\subsection{Workflow Detail}
\label{subsec:workflow}

Now, we specify the workflow details that have not yet been discussed above. 
\iffalse
To be specific, we 
first elaborate how \ptrix initializes the workflow of fuzz 
testing. Then, we describe how \ptrix ensures the correctness of the workflow 
specified in the section above.
\fi

% Following the brief introduction above, we now specify more details about the
% workflow of \ptrix. Given that \ptrix needs to initialize the key components --
% depicted in Figure~\ref{} -- prior to performing fuzz testing against a target
% program, we first elaborate the workflow of how \ptrix initializes the workflow
% of fuzz testing. Then, we describe how \ptrix ensures the correctness of the
% workflow specified in the section above.

\subsubsection{Initializing Fuzz Testing Workflow}
First, \ptrix mounts the PT module and sets it to listen to a {\tt netlink} channel.
Second, \ptrix starts the fuzzer component, which forks a child process running as the proxy
seating between the fuzzer and the target program. By passing the information
pertaining to a fuzzing task to the proxy, \ptrix triggers the proxy to
send a notification to the PT module through the established netlink channel.

On receiving the above notification, the PT module allocates a
buffer for storing PT data packets. In addition, it instantiates a variable {\tt
pt\_off} and uses it to indicate the offset of the buffer, from which to the
head of the buffer is the space where the data packets are stored. In this work, we
design \ptrix to map the buffer and the variable into the user-space of the
proxy process. In this way, we can ensure that the proxy process can retrieve
data packets without crossing the user-kernel privilege boundary, making the
performance overhead minimal.

After the PT module initialization, the proxy receives
a confirmation and further performs the following operations. First, it 
forks a child process running as the fork server. Second, the proxy process 
notices the fuzzer to generate an input and passes it to the fork server to start execution.

\subsubsection{Enforcing the Correctness of the Workflow} 
With the completion of the initialization above, \ptrix can perform fuzz testing
by following the workflow specified in Section~\ref{subsec:overview}. However, a simple
design of this workflow could potentially incur an incorrect synchronization
issue, particularly given the situation where the fuzzer, proxy, PT module and
fork server components all run concurrently. To ensure the
correctness of fuzz testing, we augment \ptrix with three callbacks 
planted into the tracepoints inside three kernel events
-- {\tt fork}, {\tt context\_switch}, and {\tt exit}. Note that 
we use the tracepoints instead of explicit interactions (such as system calls)
to avoid additional communication costs. In the following, we specify the functionality of each of these callbacks.

% The correctness of the fuzzing logic is contingent on the synchronization 
% among the concurrently running components -- Fuzzer, Proxy, PT Module and Fork Server. 
% While challenging, this brings us the opportunity to exploit the parallelism.
% %To fulfill the first condition, our design ensures \ptrix follow 
% %the workflow of \afl (as illustrated in Figure~\ref{figs:overview}). 
% Figure~\ref{figs:timeline} shows the chronological order of each step we show in Figure~\ref{figs:overview}. 
% The synchronization among different components is centered around the execution status of the target process. 
% %At the beginning of a new fuzzing run, the fuzzer prepares the input and sends it to the proxy (\textcircled{1} \textcircled{2}).
% %We reuse the design of \afl's fuzzer for this part. 
% %The proxy wakes up on receiving the input and queries its scheduler. 
% %To be specific, the scheduler checks if the previous fuzzing run has completed. If so, the scheduler permits the new run. 
% %The proxy then sends the input to the fork server (\textcircled{4}) and blocks until the target process starts. 

%The fork server then attempts to fork a child process to execute the arrived input (\textcircled{5}). 

\noindent{\bf Fork callback.}
\ptrix uses PT module to monitor the process of a target program (for brevity target process) 
and the proxy component to coordinate the entire fuzzing test. 
To facilitate this,  we introduce a {\tt fork} callback.
On the one hand, when the fork server forks the target process~\footnote{More precisely, the target process means the master thread}, 
this callback registers the target process to the proxy and 
makes the PT module ready for tracing. As such, we can ensure that
the target process does not execute until the proxy is ready and the PT module
is set up. On the other hands, this callback captures child threads forked by the 
target process, prepares these threads with the aforementioned initialization we perform to the
target process, and ensures the synchronization before these threads start.   
By doing so, \ptrix can handle multi-threading programs.  

\noindent{\bf Context\_switch callback.}
When the target process enters execution status, the CPU might switch it in and
out periodically and a {\tt context\_switch} event would occur. 
In the {\tt context\_switch}, we introduce a callback for two reasons.
 First, we design the callback to enable Intel PT to
trace a CPU core whenever the target process switches into it, and disable the
tracing at the time when the target process is switched out. In addition, this callback 
updates {\tt pt\_off} when the target process is switched out. In this way,
we guarantee that PT always writes to the right place.
Second, as PT cannot separate the traces from different threads, 
we use this callback to distinguish the target process and its child threads.
More specifically, this callback sets up PT to write in the buffer associated with 
a thread when this thread is switched in and updates the corresponding {\tt pt\_off} when 
this thread is switched out.  

% The callback of this event enables PT to trace the CPU core where the target process will be switched into. 
% To ensure that PT writes packets continuously to the PT buffer across different context switches, 
% the callback also sets PT to always start writing to the PT buffer at the offset {\tt pt\_off}. 
% Similarly, if the target process is being switched off a CPU core, 
% the callback disables PT on the core and updates the corresponding {\tt pt\_off} 
% with the ending position that PT writes to the PT buffer. 

\noindent{\bf Exit callback.}
After the target process terminates, an {\tt exit} event would occur. To use it as
a signal for concluding one round of fuzz testing, we introduce a callback in {\tt exit}. 
This callback is responsible for coordination among the fuzzer, PT module and proxy components. 
To be specific, whenever the callback is triggered, \ptrix first disables Intel PT. Then, it 
examines whether the data packets have been processed 
completely. Only with the confirmation of data packet processing completion, 
\ptrix further resets the PT module, coordinates with the fuzzer to compare the bookkeeping bitmaps and thus concludes 
one round of fuzzing testing. With this callback, we can ensure that the fuzzer 
does not conclude fuzz testing prior to the packet parsing and local bitmap 
computation. 

\subsection{Efficiency Improvement}
\label{subsec:efficiency}

To illustrate the coordination and synchronization enforced through the
aforementioned callbacks, we present the chronological order of each
component in Figure~\ref{figs:timeline}. As we can observe from the figure,
parsing data packets and computing local bitmaps sit on the critical path of
each round of fuzz testing. If these operations start after the termination of
the target process, or launch nearly simultaneously with the target process but
take a significant amount of time to complete, the fuzzing efficiency would be
significantly jeopardized and these operations would become the performance
bottleneck for \ptrix. To avoid these situations and improve performance, 
we propose a parallel and elastic PT parsing scheme and a new PT-friendly 
feedback.

\subsubsection{Parallel and Elastic PT Parsing}
\label{sec:challenge1}
%As presented above, when the target process is forked, 
%the proxy gets ready to do parsing and updating. 
%An easy design is to make the proxy wait 
%until exit of the target process and then do 
%the parsing and decoding. 

As is mentioned above, it obviously increases the time needed for a single round
of fuzz testing if \ptrix parses data packets right after the termination of a
target process. As a result, we carefully design the following scheme to perform
data packet decoding simultaneously with the target process execution.

After starting a target process, the proxy process creates a parser thread to
decode the data packets recorded through Intel PT. Depending upon how fast the
data packets are yielded, the parser thread adjusts its working status. For
example, if the parser exhausts the packets quicker than they are
recorded, it would enter an idle state until new data packets become available.
In the process of parsing data packets, we design \ptrix to maintain a variable
{\tt last\_off}, indicating the ending position where the parser thread
completes packet decoding last time. With this variable, the parser
could easily pinpoint the offset from which it could retrieve the data packets
while it is awakened from an ideal state. 

In our design, \ptrix initializes the {\tt last\_off} variable with zero. Every
time when {\tt last\_off} is less than {\tt pt\_off} -- the variable indicating
the end of the buffer that stores data packets -- the parser thread could
decode data packets and update {\tt last\_off} accordingly. With this, we can
ensure that the parser can always correctly identify the packets that
have not yet been decoded and, more importantly, guarantee that the parser 
does not retrieve data packets out of the boundary. In addition, with the 
facilitation from the {\tt exit} callback, \ptrix can ensure all data packets are processed 
behind the termination of the target process. It
should be noted that we design \ptrix to maintain these variables on the basis
of each individual thread for the simple reason that this could allow \ptrix to
handle  multi-threading.

% In this process, the parser maintains a variable {\tt last\_off} to record the offset in the PT buffer it has processed so far. This variable is 
% set to 0 initially. 
% When awaken, the parser reads {\tt pt\_off}, a variable updated by PT indicating the end of the generated trace data. 

% If 
% {\tt pt\_off} is larger than {\tt last\_off}, it means the region {\tt [last\_off, pt\_off)} is new and ready to be parsed.
% the parser then starts processing the region and updates {\tt pt\_off} with {\tt last\_off} when finished.
% This way, we ensure that the parser always correctly decodes the valid data and never read over bounds. Also note that, these variables are maintained per-thread, so as to handle multi-threading targets.

% The parser is dismissed when the target process exits. It then makes one last check to make sure that {\tt last\_off = pt\_off}. 
% At this point, the whole tracing and decoding process is complete. 

While the aforementioned design is intuitive, it is still challenging to follow
the design and perform data parsing simultaneously with the execution of a
target program. The reason is that, in order to perform data packet decoding and execute the
target process in parallel, we have to design \ptrix to update the variable {\tt
pt\_off} significantly frequently. However, due to the limitation imposed by
hardware, we can update the variable {\tt pt\_off} only at the time when a CPU
core switches out the target process. This is simply because a correct offset
can be reliably obtained only when PT tracing is disabled. In practice, our
observation, however, indicates that context switch does not frequently occur and,
oftentimes, a target process completes one round of fuzz testing without
experiencing context switch. As a result, it is infeasible to perform
simultaneous data packet parsing without disrupting the execution of the target
program.

% The above parallel design looks straightforward, but applying the design to achieve a high degree of parallelism 
% is very challenging. As we can see from the above scheme, for a high degree of parallelism, the key is to 
% update {\tt pt\_off} as frequently as possible. However, due to the limitation imposed by hardware,
% {\tt pt\_off} only gets updated when the target process is switched out from a CPU core. 
% The reason is that the correct offset can only be reliably obtained when PT tracing is disabled. To guarantee that we do not miss any trace data after disabling tracing. The most non-disrupted way is to do that when the target process is switched off the core. Nevertheless, based on our evaluation,  context switch is not sufficiently frequent. Oftentimes, one fuzzing run can finish in a single time slice (from switch-in to switch-out). Meanwhile, syscall based scheme is target-specific. Consequently, achieving parallel parsing without disrupting the target execution is challenging. 

To address the challenge above, we introduce an elastic scheme, which
leverages a timer mechanism provided by kernel to adjust the frequency of
disabling process tracing in an automated fashion. To be more specific, we first
attach a timer to a CPU core that ties to a target process. Then, we register
a handler to that timer. With this, process tracing can be enabled or
disabled, and the variable {\tt pt\_off} can be updated. For example, whenever
the timer alarm is triggered, the handler could disable the tracing, update {\tt
pt\_off} and set up the timer to arm for the next shot.

To determine the countdown for the next  alarm, we measure the length
of the data packets by retrieving the value held in the variable {\tt pt\_off}.
Then, we compare it with the variable {\tt pt\_last}, indicating the length of
the data packets that have been correctly decoded by the parser thread. Since
the value difference in these variables demonstrates the amount of data packets 
that have not yet been parsed, which reflects the speed of the parser thread
in decoding the packets. We set up the next timer alarm in an
elastic manner based on the following criteria. If the amount of the data
packets left behind exceeds a certain threshold, \ptrix decreases the countdown so that parser's workload will be reduced. Otherwise, the countdown is incremented and thus
ensuring that parser has sufficient packets to perform decoding. In
Section~\ref{sec:eval}, we demonstrate the efficiency gain obtained from this
elastic scheme by comparing it with a naive  scheme in which the parsing process
starts after the execution termination of the target process.

\begin{algorithm}[!t]
\scriptsize
\caption{Bit map updating algorithm}
\label{algo:mapupdate}
\begin{algorithmic}[1]
\INPUT  
\Statex {$trace\_bits[]$ - The bit map}
\Statex {$packet\_queue$ - The queue of PT packets}

\OUTPUT 
\Statex{Updated $trace\_bits[]$}

\Procedure{UpdateTracebits}{}
\State $bit\_hash = 0$
\State $tip\_cnt = 0$
\State $tnt\_cnt = 0$

\While{$packet\_queue.size()$}
	\State $packet$ = $packet\_queue.pop()$ 
	\If{$packet.type$ == {\tt TIP}}
		\State $bit\_hash$ = $UpdateHash(bit\_hash, packet)$
		\State $tip\_cnt$++
		\State $tnt\_cnt$ = $0$
		\If{$tip\_cnt \ge$ {\tt MAX\_TIP}}
			\State $index$ = $Encoding(bit\_hash)$
			\State $setbit(trace\_bits, index)$
			\State $tip\_cnt$ = $0$
			\State $bit\_hash$ = $UpdateHash(0, packet)$ \Comment{Start a new slice}
		\EndIf 
	\EndIf
	\If {$packet.type$ == {\tt TNT} {\tt \&\&} $tnt\_cnt \le$ {\tt MAX\_TNT}}	
		\State $bit\_hash$ = $UpdateHash(bit\_hash, packet)$
		\State $tnt\_cnt$++
	\EndIf	
\EndWhile
\EndProcedure
\end{algorithmic}
\end{algorithm}

%\begin{algorithm}[t]
%\scriptsize
%\caption{Hash algorithm}
%\label{algo:hash}
%\begin{algorithmic}[1]
%\INPUT  
%\Statex {$bit\_hash$ - 64-bit hash value of previous packets}
%\Statex {$packet$ - A new PT packet}
%
%\OUTPUT
%\Statex{$hash$ - Result of new hash value}

%\Procedure{UpdateHash}{}
%\State $hash$ = $bit\_hash$
%\For{$byte$ {\bf in} $packet.data$} 
%	\State $hash$ = $byte$ + $hash$ \verb|<<| $6$ + ($hash$ \verb|<<| $16$) - $hash$
%\EndFor
%\State \Return $hash$
%\EndProcedure
%\end{algorithmic}
%\end{algorithm}
\begin{algorithm}[t]
\scriptsize
\caption{Encoding algorithm}
\label{algo:encoding}
\begin{algorithmic}[1]
\INPUT  
\Statex {$bit\_hash$ - A 64 bit hash value to encode }
\OUTPUT
\Statex{$index$ - Result of the encoding}

\Procedure{Encoding}{}
\State $bit\_size$ = $bit\_map.size$ \verb|<<| 3 \Comment{Number of bits in $bit\_map$}
\State $range$ = {\tt U64\_MAX} \verb|>>| (64 - $\log_2(bit\_size)$)
\State $rnd$ = 64 / $bit\_size$
\State $index$ = $bit\_hash$ \& $range$ 
\For{ $k$ $\leftarrow$ 0 {\bf to} $rnd$}
	\State	$bit\_hash$ = $bit\_hash$ \verb|>>| $bit\_size$
	\State $index$	$\oplus$ = $bit\_hash$ \& $range$

\EndFor
\State \Return $index$ \& $range$
\EndProcedure
\end{algorithmic}
\end{algorithm}

\subsubsection{New PT-friendly Feedback Scheme}
As is mentioned earlier, if parsing data packets incurs significant latency, the
improvement in fuzzing efficiency obtained from the aforementioned parallel parsing scheme
would become a futile attempt. Therefore, in addition to taking
advantage of parallelization for improving the efficiency of fuzz testing, we
need an efficient approach to decode data packets and thus expedite each round
of fuzz testing.

Intuitively, we can perform data packet decoding by following the footprints of
previous works~\cite{winafl, kafl}, in which fuzzing tools are designed to
reconstruct instructions executed -- using the technique discussed in
Section~\ref{sec:bg} -- and then compute the bitmaps by following the bitmap
update algorithm introduced by \afl. However, as we demonstrated in
Section~\ref{sec:eval}, using such an approach, dubbed {\tt Edge-PT}, as part of our fuzzing system 
does not actually introduce any efficiency improvement. This is simply because recovering instructions from data packets incurs a significant amount of latency even when we cache the disassembling results. 

%%%%%%%%%%%%%%%%%%%%%%%%%%%%%%%%%%%%%%%%%%%%%%%%%%%%%%%%%%%%%%%%%%%%%%%%%%%%%%%
% 
% [comment from xxy]
% Add a table here to indicate the algorithm of how to do hash to bitmap 
% projection 
% 
%%%%%%%%%%%%%%%%%%%%%%%%%%%%%%%%%%%%%%%%%%%%%%%%%%%%%%%%%%%%%%%%%%%%%%%%%%%%%%%

To address this issue, we introduce a new scheme
 to compute and update bitmaps needed for fuzz testing. 
At the high level, we concatenate the  {\tt TIP} and {\tt TNT} packets 
into ``strings'' and then hash those strings into indices for bitmap updating. 
We describe the details of our algorithm as follows. 

The overall algorithm is presented in Algorithm~\ref{algo:mapupdate}. 
As fuzzing continues, the {\tt UPDATETRACEBITS} procedure consumes 
the {\tt TIP} and {\tt TNT} packets produced by our decoder. Note that 
for better efficiency, single-bit {\tt TNT}s are concatenated 
into byte-aligned packets. In {\tt UPDATETRACEBITS}, 
each packet is taken by the {\tt UpdateHash}
routine to update a hash value. 
%which is {\tt 0} initially (Algorithm~\ref{algo:hash}).  
{\tt UpdateHash} implements the SDBM hash function which supports 
streaming data~\cite{seltzer1991new}. We selected SDBM because it 
has been demonstrating great over-all distribution for various data 
sets~\cite{henke2008empirical} and it has low computation complexity.

When {\tt UPDATETRACEBITS} sees MAX\_TIP {\tt TIP} packets, 
it encodes the accumulated hash value as an index to update the bitmap.
This essentially cuts the packets stream into slices and record each 
slice with a bit. Shortly we will explain the rationale behind this design and 
how we determine MAX\_TIP. Encoding of the hash value is achieved using Algorithm~\ref{algo:encoding}.
It transforms a 64-bit hash to a value in $[0, max\_bit\_index)$, where $max\_bit\_index$ represents the number of bits in the bitmap. 
To be more specific, this encoding splits the hash value into multiple pieces 
with each piece converged into $[0, max\_bit\_index)$. Then it exclusively-ors 
these pieces to form the index. Given a set of equally distributed hash values, 
Algorithm~\ref{algo:encoding} will ensure that they
are mapped into $[0, max\_bit\_index)$ with uniform distribution.  

In this design, we only spare one bit to record the appearance of a slice. 
This differs from the design of {\tt AFL} --- {\tt AFL} uses one byte to 
log not only the appearance of an edge but also its hit count. Our design 
is motivated by the observation that most of the slices (under the MAX\_TIP
we select) only arise once, which only require single bits for recording.  
As a result, our scheme uses {$7$}x less space than the hit-count-recording 
scheme in {\tt AFL}. This, in turn, enables {\tt bit\_map} to better reside in 
{\tt L1} cache. As we will show in Section~\ref{sec:eval}, this choice brings
around an additional 8\% speed up.

The above algorithm avoids the expensive re-construction of instruction trace.  
As we will shortly show in Section~\ref{sec:eval}, it brings us over 10x acceleration on execution speed.
Essentially, this algorithm alters {\tt AFL}'s code-coverage based feedback in {\tt AFL} to ``control-flow'' based feedback.   
In the following, we discuss how our design maintains the functionality and gains the efficiency. 

Functionality wise, our new feedback provides guidance that is inclusive of code coverage 
(the feedback natively used by \afl). The guidance requires that the feedback to diverge 
when inputs incur different execution behaviors. The feedback to guide {\tt AFL}
captures new code edges and their new hitting counts. 
Going beyond {\tt AFL}, our feedback acutally approaches a higher 
level of guidance --- \emph{path guidance}. More specifically, our feedback encodes 
the control flow packets, which uniquely represents an execution path. 
Following inputs that lead to different execution paths, our feedback produces different outputs.
Therefore, it captures not only new code edges and new hitting counts of code edges, 
but also new orders and new combinations among code edges, since all the four events result in 
new execution paths.

\begin{figure}[t]
    \centering
    \begin{subfigure}[t]{0.42\textwidth}
        \centering
        \includegraphics[scale=0.8]{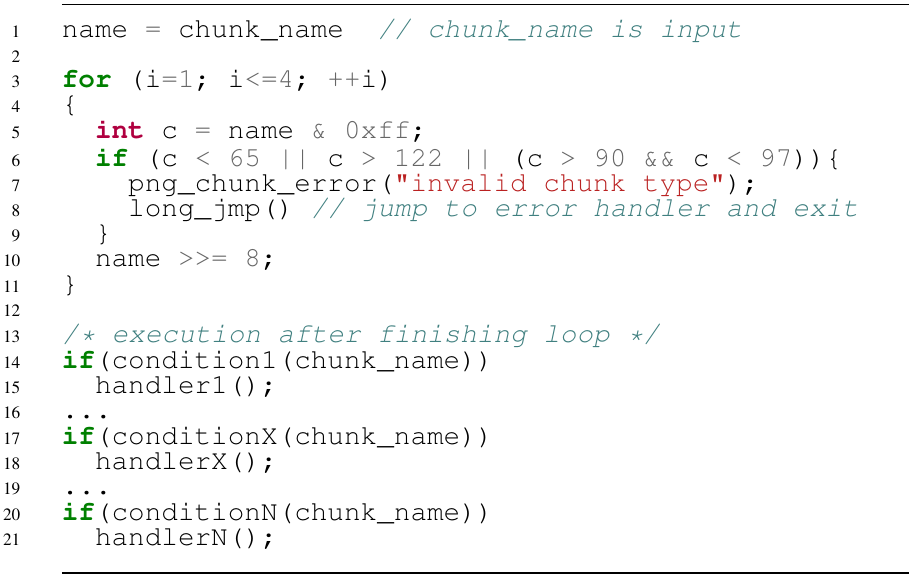}
        \caption{A code fragment in {\tt libpng-1.6.31}. \ptrix can generate inputs to reach handlerX while {\tt AFL} could not.}
	\label{code:libpng}
    \end{subfigure} \\
    \begin{subfigure}[t]{0.42\textwidth}
        \centering
        \includegraphics[scale=0.32]{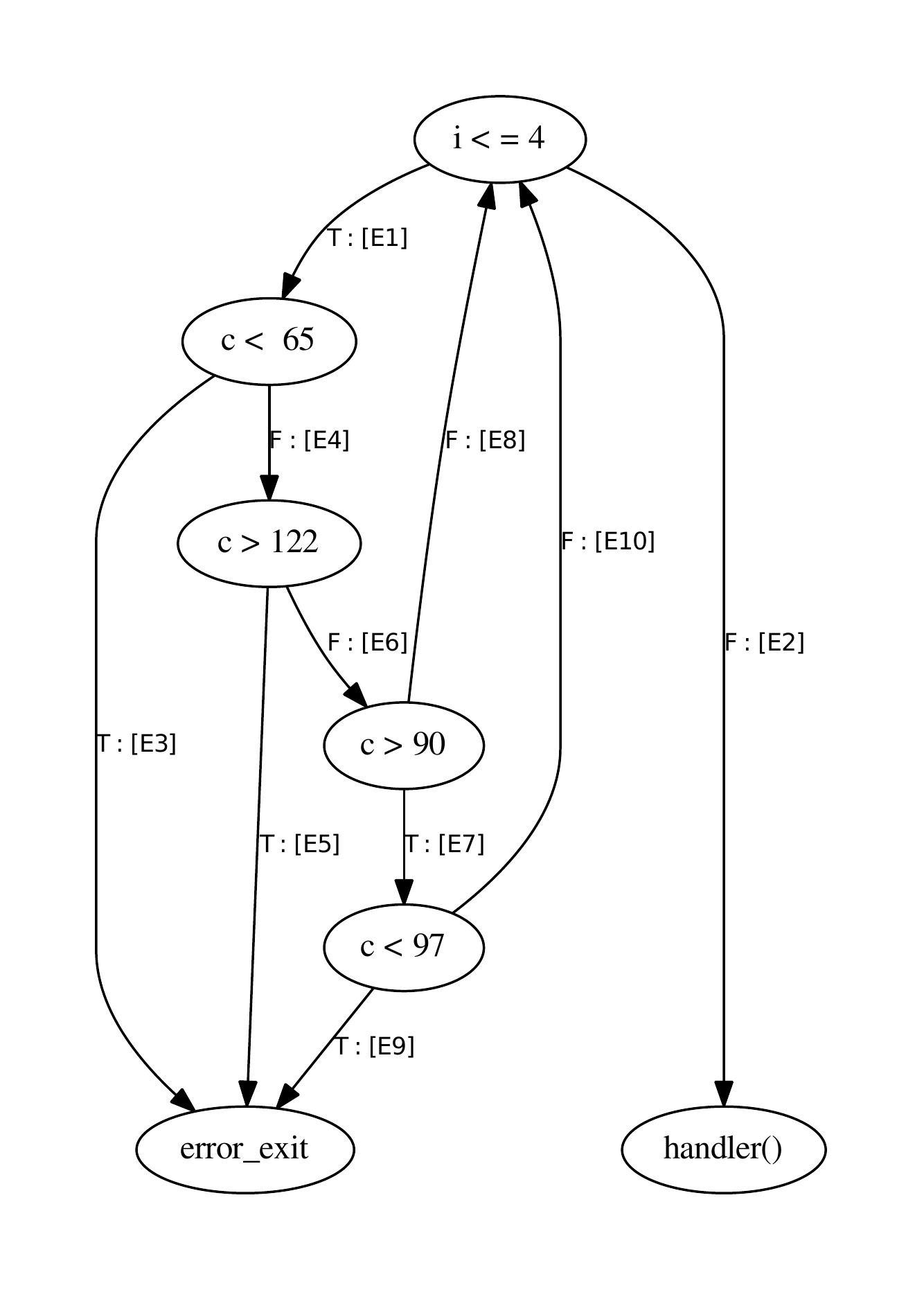}
        \caption{Control flow graph of the code shown above. On an edge, ``T/F'' means true/false and ``[EX]'' is the number of the edge.}
	\label{code:cfg}
    \end{subfigure}
    \caption{An example for new code coverage by \ptrix}
    \vspace{-2em}
\end{figure}

Efficiency wide, our new feedback may encounter two caveats when mounted for fuzzing. In the following, 
we introduce their details and explain our solutions. 

First, we need a giant bitmap to record the tremendous volume of distinct execution paths.  
This greatly impacts the frequent bitmap updating and comparison,
mainly because of a reduced cache hit ratio and increased comparing operations.  
To mitigate this, we split an entire path into slices aligned by MAX\_TIP {\tt TIP}s.
The rationale behind is that a smaller MAX\_TIP reduces the size of a slice, which consequently shrinks 
the permutation space of slices and the needed bitmap. However,  
intuition suggests that decreasing MAX\_TIP will also reduce the path guidance.  
To better balance the efficiency and guidance, we pick MAX\_TIP following two criteria: 
(1) \ptrix works with a 64KB bitmap~\footnote{The 64KB bitmap is used in {\tt AFL} by default} 
--- We confirm this if the bitmap increases no faster than 30\% per 24 hours; 
(2) \ptrix achieves an equivalent (if not better) guidance than {\tt AFL} --- We confirm this if 
\ptrix rarely runs into collisions using a corpus of seeds generated by {\tt AFL} in 24 hours.  

Second, our new feedback may cause \ptrix to overly explore or even 
get trapped in localized code segments (in particular loops), which slows down 
or impedes \ptrix to explore new code. We invest two-fold efforts in addressing this issue. 
At first, we restrict the number of {\tt TNT}s between two {\tt TIP}s (line {\tt 18} in Algorithm~\ref{algo:mapupdate}) 
and we call this approach \emph{descending path guidance}.
Our reason is that extremely long TNT sequences
are typically due to massive iterations in loops. Limiting the number
of {\tt TNT}s can effectively prevent \ptrix from trapping into deep loops. 
Note that determination of MAX\_TNT is explained in Section~\ref{sec:impl}. 
In addition, we adjust the fuzzing scheduling in \ptrix to prefer 
seeds producing small {\tt TNT} sequences.  As we will show in Section~\ref{sec:eval}, the two ideas
well avoid over-exploration of localized code and enable \ptrix 
to achieve high fuzzing efficiency. 

\subsection{Side Benefits of New Feedback Scheme}
\label{sec:libpng}
Our testing with \ptrix on benchmark programs illustrates that 
the newly designed feedback truly has stronger guidance which brings side benefits 
to fuzzing.

\noindent{\bf Better code coverage.}
Recall that our feedback has the advantage of capturing the orders and combinations of traversed code edges. 
This property benefits \ptrix in covering code that \afl is unable to reach.  Figure~\ref{code:libpng}
showcases such an example in {\tt libpng-1.6.31}. The code verifies a 4-byte field  {\tt chunk\_name} 
in the image header through a loop (line {\tt 3 - 11}). Any one of the four bytes violating the checks (line {\tt 6}) will break 
the loop and result in an early exit (line {\tt 8}). A valid {\tt chunk\_name}, will be processed by a handler corresponding 
to its type (line {\tt 14 - 21}).

In the fuzz testing, \afl generated inputs whose first byte violated the three checks in different ways (line {\tt 7}). 
These inputs followed different execution paths (as shown in Figure~\ref{code:cfg}), including 
\{{\tt E1 $\to$ E3}\}, \{{\tt E1 $\rightarrow$ E4 $\rightarrow$ E5}\}, 
\{{\tt E1 $\rightarrow$ E4 $\rightarrow$ E6 $\rightarrow$ E7 $\rightarrow$ E9}\}. 
By further mutating those inputs, \afl produced test cases which 
chronologically explored the following paths: \{{\tt E1 $\rightarrow$ E4 $\rightarrow$ E6 $\rightarrow$ E8 $\rightarrow$ E1 $\rightarrow$ E3}\}, 
\{{\tt E1 $\rightarrow$ E4 $\rightarrow$ E6 $\rightarrow$ E8 $\rightarrow$ E1 $\rightarrow$ E4 $\rightarrow$ E6 $\rightarrow$ E8 $\rightarrow$ E1 $\rightarrow$ E3}\} 
and \{{\tt E1 $\rightarrow$ E4 $\rightarrow$ E6 $\rightarrow$ E8 $\rightarrow$ E1 $\rightarrow$ E4 $\rightarrow$ E5}\}. 
As the third test case led to neither new edge nor new hitting count, it was ignored by \afl. 
But in fact, permutating this input would result in a valid {\tt chunk\_name}, 
which matches {\tt conditionX} and makes {\tt handlerX} executed. 
Different from \afl, \ptrix values this discarded input since it triggered new combinations of code edges, which 
enables \ptrix to ultimately reach {\tt conditionX}. 

\begin{figure}[t]
\includegraphics[scale=0.8]{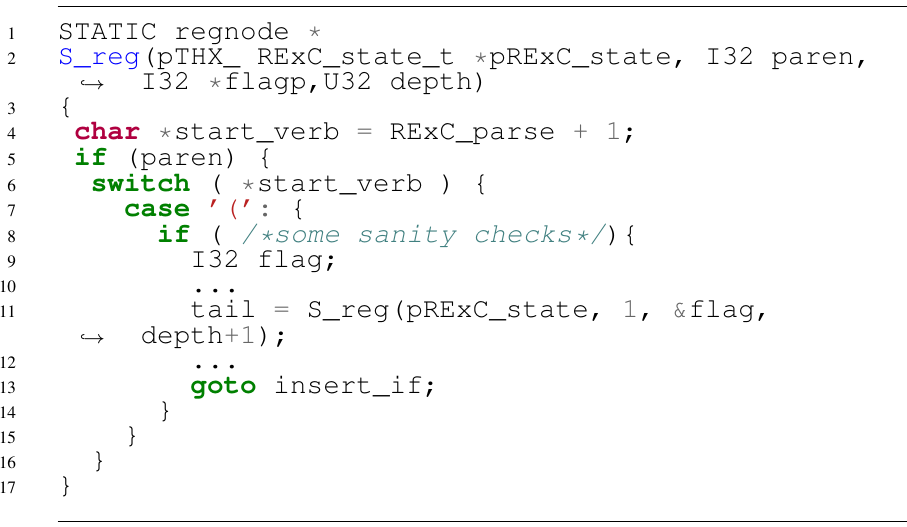}
\caption{A stack exhaustion bug in Perl. }
\label{code:perl_bug}
\vspace{-2em}
\end{figure}

\noindent{\bf Uncover deep bugs.}
As our new feedback provides additional guidance, 
 \ptrix explores code segments more comprehensively, 
leading to coverage of deeper execution space. This helps the discovery of not only new code 
space but also deeper defects. In Figure~\ref{code:perl_bug}, 
we demonstrate such a case \ptrix identified in {\tt perl-5.26.1}.
With an input containing over {\tt 3500} ``('', one can trigger a stack exhaustion error. 
Specifically, each of those ``('' would trigger a recursive call to function {\tt S\_reg} at line {\tt 11},
which gradually exhausted the stack region. 

\afl records the feedback pertaining to an edge with a single byte, which may log at most 255 hits.
As such, \afl ignores inputs that invokes more than 255 recursions.  
This prevents \afl from mutating those inputs towards chasing down the bug. 
While in \ptrix, deeper recursions produce new {\tt TNT} sequences, which is captured by the new feedback.

\section{Implementation}
\label{sec:impl}

We implemented \ptrix on 64-bit Ubuntu 14.04-LTS 
and released our prototype implementation at \url{https://github.com/junxzm1990/afl-pt}. We tested \ptrix on a set of machines armed with various
Intel processors, including Core i7-6770HQ, Core i7-6700K Skylake-H series and
Core i5-7260U Kaby Lake series. In the following, we highlight the important
implementation details.

\noindent{\bf Fuzzer.} To provide \ptrix with better usability, we integrate the main
fuzzing logic of \ptrix into the fuzzer of \afl ({\tt afl-fuzz}). In this way,
 a user of \ptrix only needs to specify a
flag ({\tt -P}) along with other options that are identical to those defined by \afl.

\noindent{\bf Proxy.} Recall that one of the major tasks for the proxy is to
parse PT packets. To obtain the optimal performance
in terms of parsing efficiency, we implement the packet parser by porting the
decoder of {\tt Griffin}~\cite{griffin}. We trim the decoder by removing the control flow reconstruction steps and add supports for our elastic decoding. Our new decoder contains less than {\tt 300} lines of code.
As is described in Section~\ref{subsec:efficiency}, the proxy component of
\ptrix sets up a threshold to restrict the number of {\tt TNT} packets in a 
sub-trace. To determine this threshold, we also implement a subroutine
for the proxy component, which utilizes {\tt angr}~\cite{angr} -- a binary
analysis tool -- to count the number of basic blocks in a target program and
then deems that number as the value of the threshold. The reason behind this is
that we observed the number of TNT packets is proportionate to the size of the traced program.

\noindent{\bf PT Module.} We developed the PT module as a separate loadable kernel
module (LKM). At the high level,  the module manages Intel PT and communicates
with the proxy component. Technically, we implement the module to enable PT to
run in the  Table of Physical Addresses (ToPA) mode. In this mode, Intel PT can
store the tracing packets in multiple discontinuous physical memory areas. 
For flexibility, the size of the overall
trace buffer can be configured via a parameter when installing the PT module.
Considering the tracing buffer could get fully occupied,  we implement the PT
module to handle that situation by clearing the {\tt END} bit and setting the
{\tt INT} bit in the last ToPA entry. By doing this, Intel PT could trigger a
performance-monitoring interrupt when the tracing buffer is fully occupied.
Since this interrupt may have a skid and result in a loss of PT packets,  we
further append an entry to the end of ToPA which also points to a {\tt 4} MB
physical memory area.

%In addition to the flexible implementation above, we introduce some
%configuration to optimize the performance of the PT module. To reduce the
%workload of the PT parser, we configure Intel PT to ignore the packets
%irrelevant to our fuzzing logic, e.g., packets for kernel-level execution and
%timing-related packets.  To eliminate unnecessary communication overhead, we map
%certain memory allocated by  the PT module to the address space of the proxy
%component, e.g., those status variables and the PT Buffer. Technically,  we
%leveraged the kernel API {\tt remap\_pfn\_range} to implement this mapping
%function.

\noindent{\bf Fork Server.} We compiled the fork server into the
{\tt GNU ld} linker and used it through a series of configurations.  During the
target program initialization, our linker gets started and completes its works
on linking and loading. It then enters the forking loop  as we described in
Section~\ref{sec:bg}. 

\section{Evaluation}
\label{sec:eval}

\definecolor{mygray}{gray}{0.9}

\begin{table}[t!]
\centering
\scriptsize
\begin{tabular}{lcc|cc}
\toprule[0.5pt]
\toprule[0.5pt]

\multicolumn{3}{c|}{\bf{\emph{Program}}} & \multicolumn{2}{c}{\bf{\emph{Settings}}}   
 \\ \hline
 {\tt Name} & {\tt Version} & {\tt Driver} & {\tt Seeds} & {\tt Options}
 \\ \hline

\rowcolor{mygray}
libpng & 1.6.31 & readpng & supplied by AFL & empty \\

libjpeg & jpeg-9b   & djpeg & supplied by AFL  & ``-gif''  \\

\rowcolor{mygray} 
libxml & 2.29  & xmllint &  supplied by AFL & empty \\

c++filt & 2.29 & cxxfilt & empty byte & empty \\

\rowcolor{mygray}
nm & 2.29 & nm-new & supplied by AFL & empty \\

objdump & 2.29 & objdump & supplied by AFL & ``-D'' \\

\rowcolor{mygray}
exif+libexif & 0.6.21 & exif & ~\cite{ianareex74:online} & empty \\

perl & 5.26.1 & perl & ~\cite{FuzzingP53:online} & empty \\

\rowcolor{mygray}
mupdf & 1.11 & mutool & supplied by AFL & ``show'' \\

\bottomrule[0.5pt]
\bottomrule[0.5pt]
\end{tabular}

\caption{Evaluation settings}
\label{tab:eval-setup}
\vspace{-2.5em}
\end{table}

In this section, we present the evaluation of \ptrix in terms of fuzzing
efficiency and vulnerability discovery.

For efficiency, we performed two sets of experiments. First, we compare \ptrix with
{\tt QEMU-AFL}, {\tt Edge-PT}, and {\tt PTFuzzer}~\cite{zhang2018ptfuzz} on execution speed. {\tt QEMU-AFL} refers to \afl running in the {\tt QEMU} mode and {\tt Edge-PT} 
is a ported version of {\tt kAFL}~\cite{kafl} that supports user space application. 
This set of experiments aims to illustrate the efficiency improvement of \ptrix on executing
the same amount of inputs. Second, following the best practise~\cite{evalfuzz}, 
we evaluated \ptrix on efficiency of code
coverage, which is a widely accepted utility metric of fuzzers~\cite{fairfuzz,
li2017steelix}. Recall that \ptrix uses feedback that has higher
path-sensitivity than {\tt QEMU-AFL}. To show that our new feedback indeed allows \ptrix
to discover new code space, we also conducted a study to compare the code space
explored by \ptrix and {\tt QEMU-AFL}.

To evaluate its vulnerability discovery ability, we applied \ptrix on a set of
commonly used and exhaustively fuzzed programs. As we will present shortly,
\ptrix discovers \numbug new vulnerabilities. Among them, at least \textbf{10} were
discovered due to our new feedback.

\subsection{Experiment Settings}

To support our evaluation, we selected a set of {\tt 9} programs. Details about
these programs and the corresponding fuzzing settings are presented in
Table~\ref{tab:eval-setup}. All these programs are either commonly used for
fuzzing evaluation~\cite{vuzzer, aflfast} or treated as core software by the
Fuzzing Project~\cite{TheFuzzi90:online}. In addition, they represent a high
level of diversity in functionality and complexity. Considering that different
seed inputs and execution options could lead to varying fuzzing
results~\cite{sage}, we used the seeds suggested by \afl and configured the
options following the existing works. 

For consistency, we conducted all the experiments on machines equipped with Intel
Core i5-7260U and {\tt 8 GB} RAM running 64-bit Ubuntu 14.04-LTS. To
minimize the effect of randomness introduced during software fuzzing,  we ran
each fuzzing test {\tt 5} times and reported the average results with standard
deviation.

\subsection{Execution Speed Evaluation}
\label{subsec:eval-speed}

\begin{figure}[h]
\includegraphics[scale=0.32]{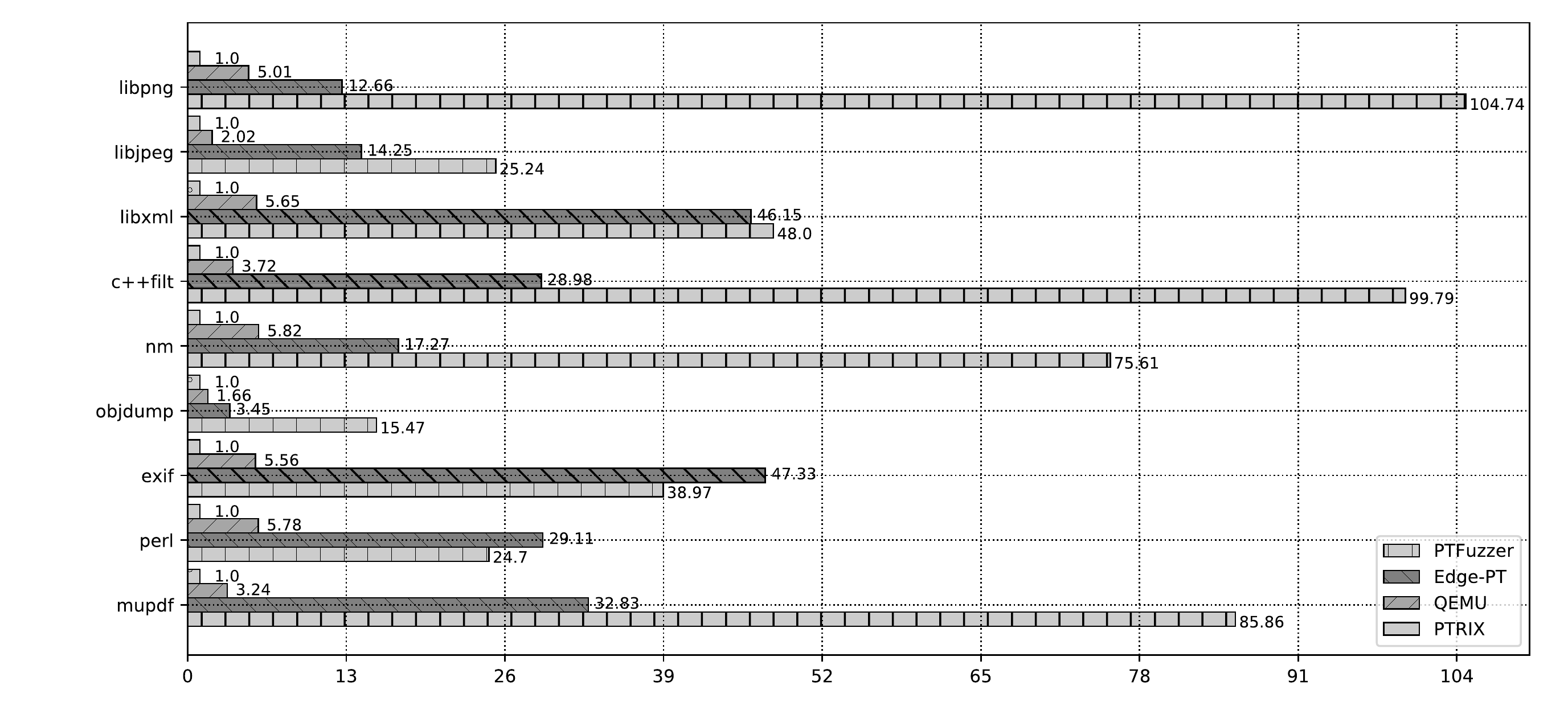}
\caption{Normalized dry-run duration for different fuzzing techniques. Shorter is better.}
\label{fig:dryrun}
\vspace{-1em}
\end{figure}

To show how fast is \ptrix, we compared its execution speed with {\tt QEMU-AFL},
{\tt Edge-PT}, and {\tt PTFuzzer}. To be specific, 
we ran these fuzzers with an identical
input corpus and examined their execution time. In this evaluation, different
inputs could trigger different types of fuzzing operations/decisions. For
example, an input that results in no new coverage will be discarded without
further processing. To avoid such difference in fuzzing runs, we selected inputs
which make all the fuzzers to go through the entire fuzzing procedure. For this,
we ran {\tt QEMU-AFL} with the settings shown in Table~\ref{tab:eval-setup} for 
{\tt 24} hours and only kept inputs that led to new coverage. 
Note that these inputs also resulted
in new coverage in \ptrix due to its highly sensitive feedback.

In this test, we utilized the \emph{dry-run} mode of \afl. It allows the fuzzers
to repeatedly process the above input corpus. In Figure~\ref{fig:dryrun}, we
show the evaluation results that have been normalized with \ptrix as baseline.
On average, \ptrix ran {\tt 4.3x}, {\tt 25.8x}, and {\tt 54.9x} faster than {\tt QEMU-AFL}, {\tt Edge-PT}, and {\tt PTFuzzer}~\footnote{Our evaluation on {\tt PTFuzzer} shows much worse performance than the results reported in~\cite{zhang2018ptfuzz}. We believe this is mainly because our benchmarks have higher complexities and the seeds we use trigger deeper execution.}, respectively. In addition, we observed that {\tt Edge-PT} ran
{\tt 6.0x} slower than {\tt QEMU-AFL} with all our optimization enabled (\ie
parallel decoding, optimized communication and caching instruction trace), 
and {\tt PTFuzzer} ran {\tt 13.5x} slower than {\tt QEMU-AFL}. 
Considering that {\tt Edge-PT}, {\tt PTFuzzer} and 
{\tt AFL-QEMU} share the identical feedback, this observation indicates that 
the design with control flow reconstruction cannot truly expose the potential of 
PT in improving fuzzing efficiency.

\begin{table}[t!]
\centering
\scriptsize
\begin{tabular}{l|ccc}
\toprule[0.5pt]
\toprule[0.5pt]

\multicolumn{1}{c|}{\bf{\emph{Program}}} & \multicolumn{3}{c}{\bf{\emph{Optimization}}}   
 \\ \hline
 {\tt Name} & \parbox[t]{2.5cm}{\centering {\tt New Feedback\\ Scheme}} & \parbox[t]{1.5cm}{\centering {\tt Parallel \\ Parsing}} & \parbox[t]{2.0cm}{\centering {\tt Bitmap \\ Optimization}}
 \\ \hline

\rowcolor{mygray}
libpng & 1038.16\% & 22.19\% & 9.51\% \\

libjpeg & 1412.96\% & 36.54\% & 14.04\% \\

\rowcolor{mygray}
libxml & 2856.08\% & 51.02\% & 5.80\% \\

c++filt & 1208.11\% & 28.70\% & 10.58\% \\

\rowcolor{mygray}
nm & 1145.90\% & 18.22\% & 6.61\% \\

objdump & 393.58\% & 6.63\% & 4.40\% \\

\rowcolor{mygray}
exif+libexif & 2695.03\% & 41.64\%  & 8.60\% \\

perl & 845.71\% & 63.14\% & 9.37\% \\

\rowcolor{mygray}
mupdf & 1426.82\% & 47.61\% & 3.68\% \\

{\bf Average} & 1446.93\% & 35.08\% & 8.07\% \\

\bottomrule[0.5pt]
\bottomrule[0.5pt]
\end{tabular}
\caption{\ptrix system optimization breakdown}
\label{tab:breadown}
\vspace{-2.5em}
%\vspace{-2ex}
\end{table}

\point{\ptrix optimization breakdown} 
\label{subsec:eval-breakdown}
To better understand how \ptrix achieves the high execution speed, we inspected
the improvement that each of our optimization introduces. We first re-ran
\ptrix without our new feedback scheme, parallel trace decoding and bitmap
optimization. Then we enabled the optimization one by one and measured the
increase of execution speed independently. The results are shown in
Table~\ref{tab:breadown}. On average, our new coverage scheme increases the
execution speed by over {\tt 14X}. The major reason, we believe, is that the new
scheme avoids the time-consuming instruction reconstruction. In addition, the
parallel parsing introduces {\tt 35\%} increase in execution speed and our
bitmap optimization contributes around {\tt 8\%} to the speedup. 

\begin{figure*}[t]
    \centering
    \begin{subfigure}[b]{0.325\textwidth}
        \centering
        \includegraphics[width=0.8\textwidth]{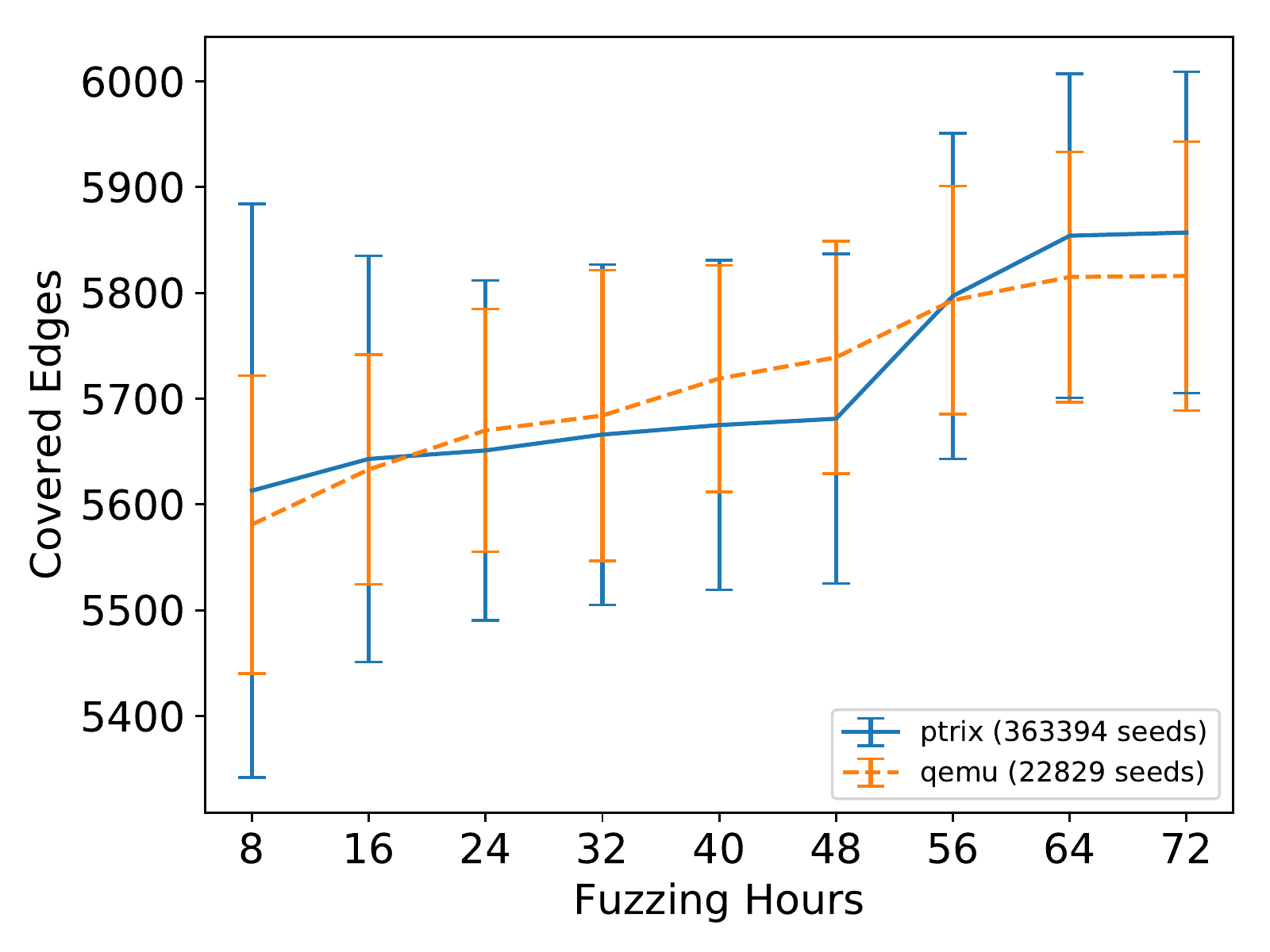}
        \caption{\scriptsize{c++filt*}}
        \label{fig:x++filt}
    \end{subfigure}
    \begin{subfigure}[b]{0.325\textwidth}
        \centering
        \includegraphics[width=0.8\textwidth]{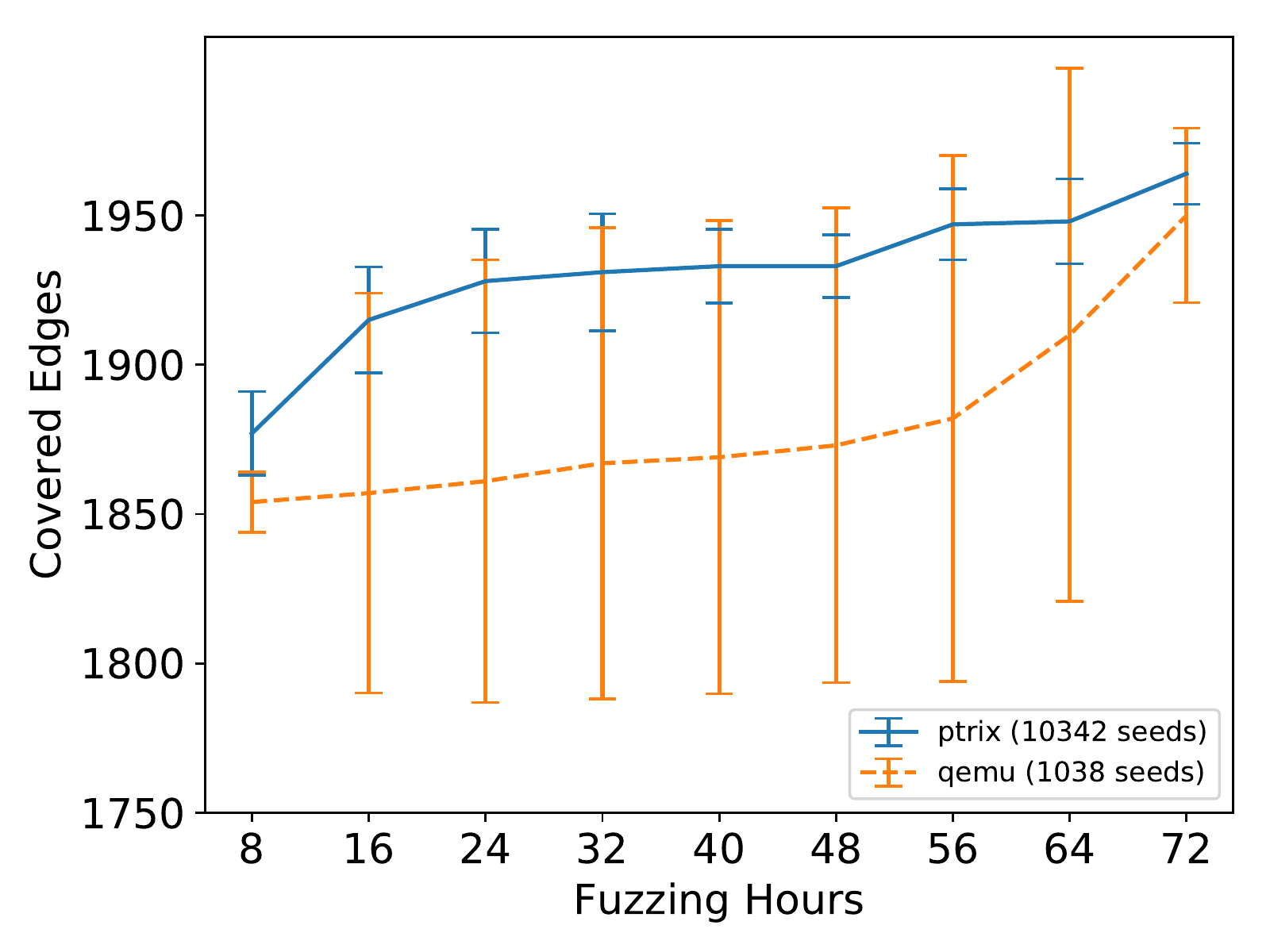}
        \caption{\scriptsize{exif}}
        \label{fig:exif}
    \end{subfigure}
    \begin{subfigure}[b]{0.325\textwidth}
        \centering
        \includegraphics[width=0.8\textwidth]{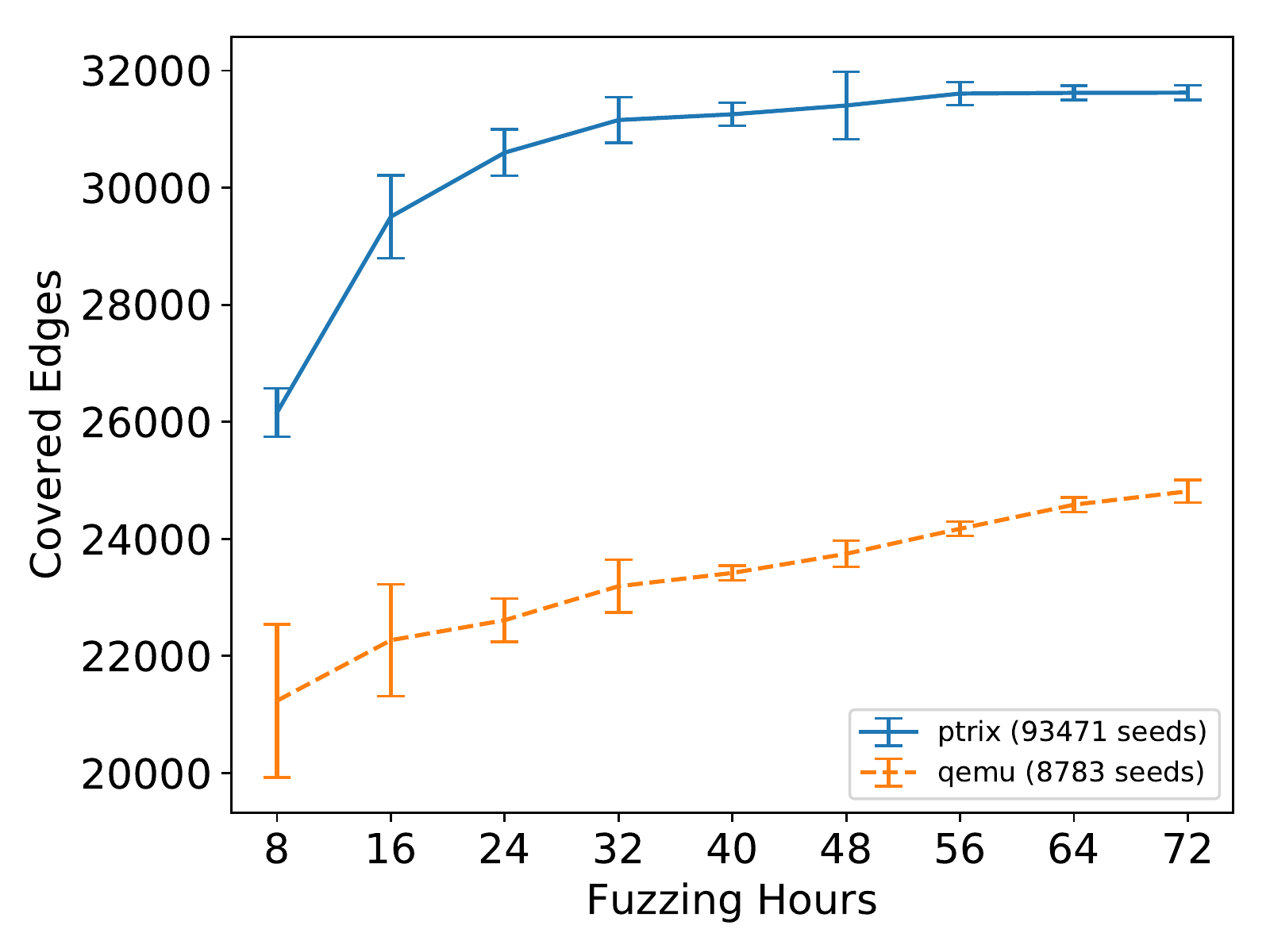}
        \caption{\scriptsize{perl}}
        \label{fig:perl}
    \end{subfigure}
    \\
    \begin{subfigure}[b]{0.325\textwidth}
        \centering
        \includegraphics[width=0.8\textwidth]{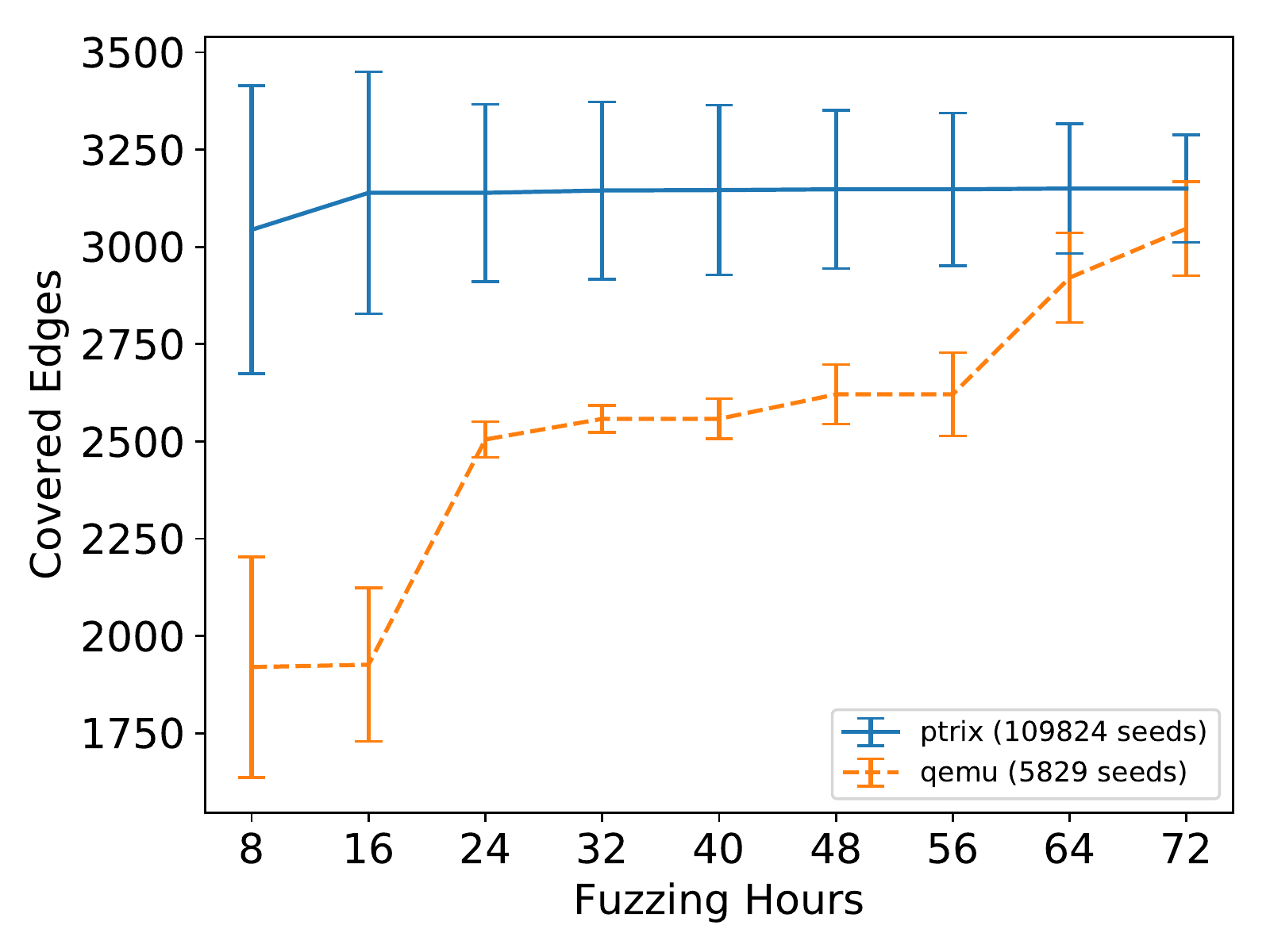}
        \caption{\scriptsize{libjpeg*}}
        \label{fig:libjpeg}
    \end{subfigure}
    \begin{subfigure}[b]{0.325\textwidth}
        \centering
        \includegraphics[width=0.8\textwidth]{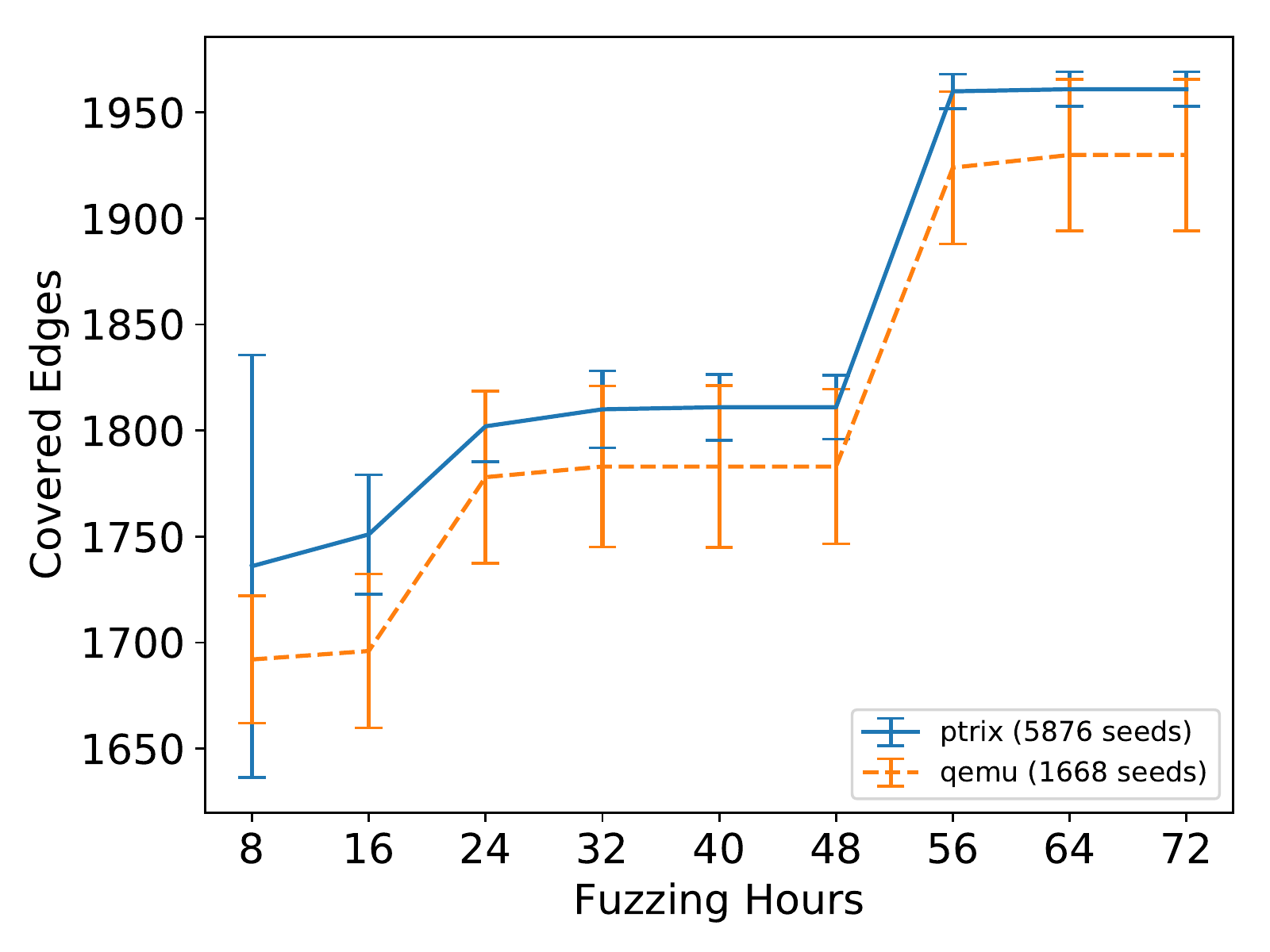}
        \caption{\scriptsize{libpng*}}
        \label{fig:libpng}
    \end{subfigure}
    \begin{subfigure}[b]{0.325\textwidth}
        \centering
        \includegraphics[width=0.8\textwidth]{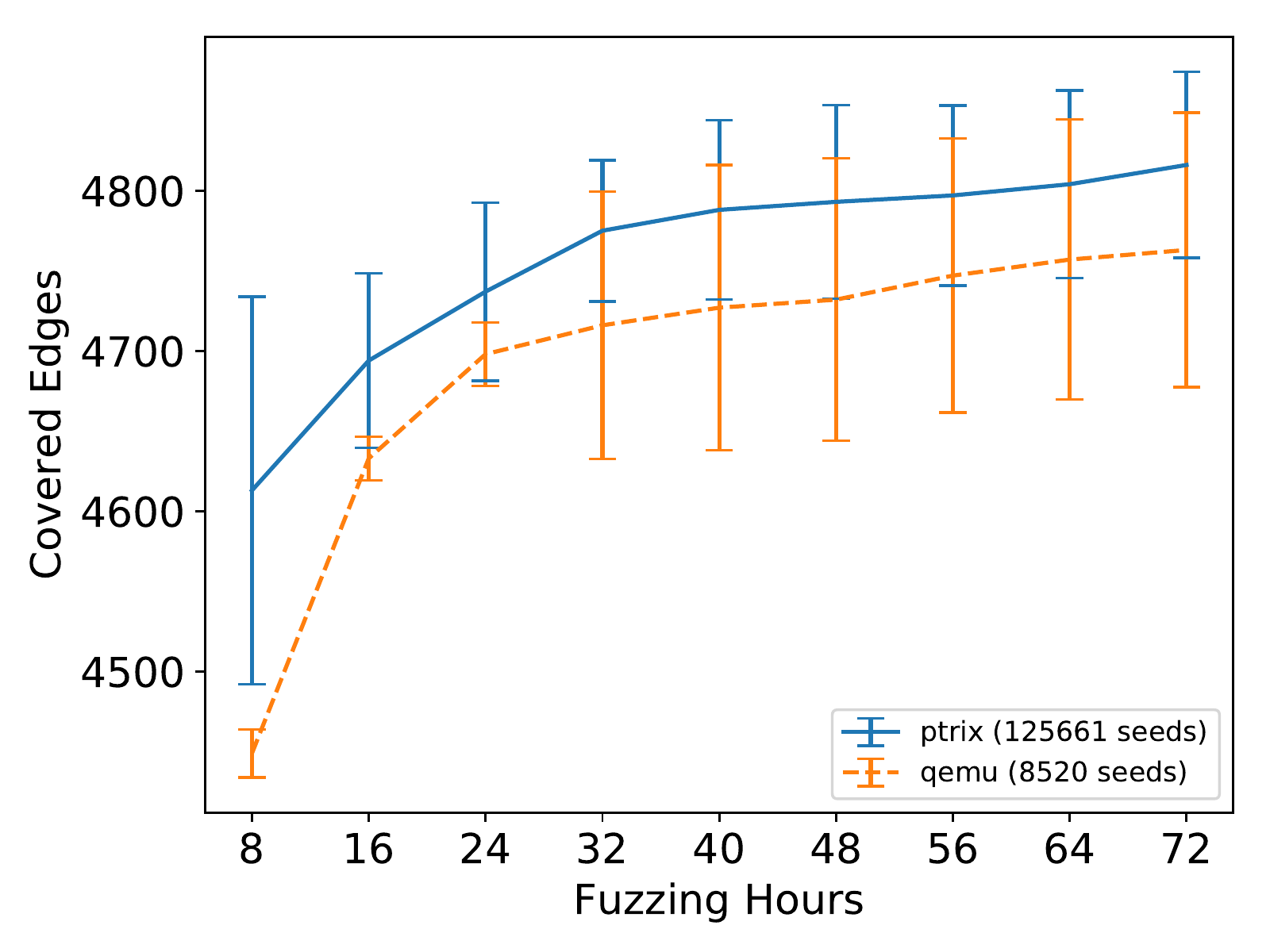}
        \caption{\scriptsize{libxml}}
        \label{fig:libxml}
    \end{subfigure}
    \\
    \begin{subfigure}[b]{0.325\textwidth}
        \centering
        \includegraphics[width=0.8\textwidth]{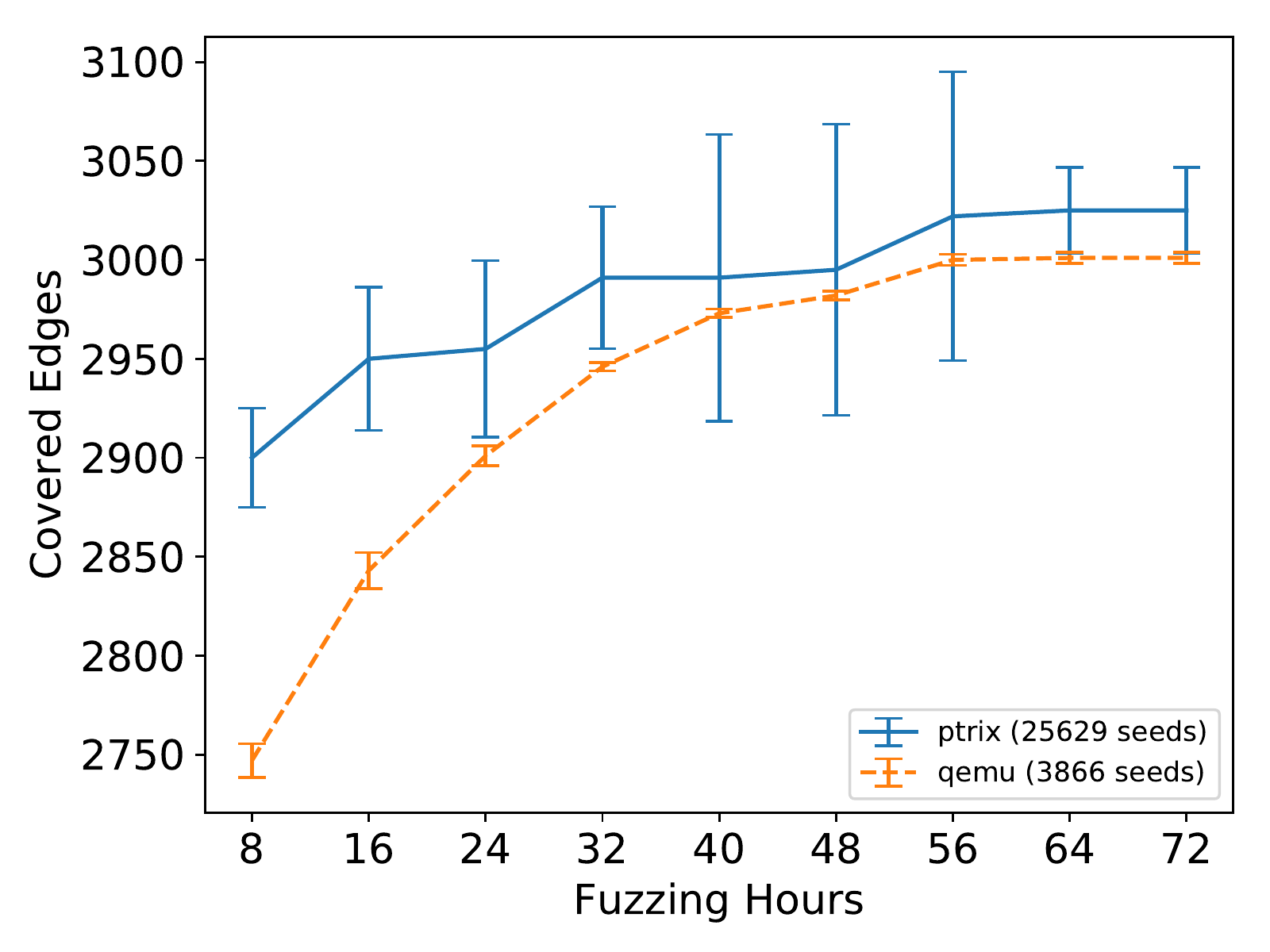}
        \caption{\scriptsize{mupdf*}}
        \label{fig:mupdf}
    \end{subfigure}
    \begin{subfigure}[b]{0.325\textwidth}
        \centering
        \includegraphics[width=0.8\textwidth]{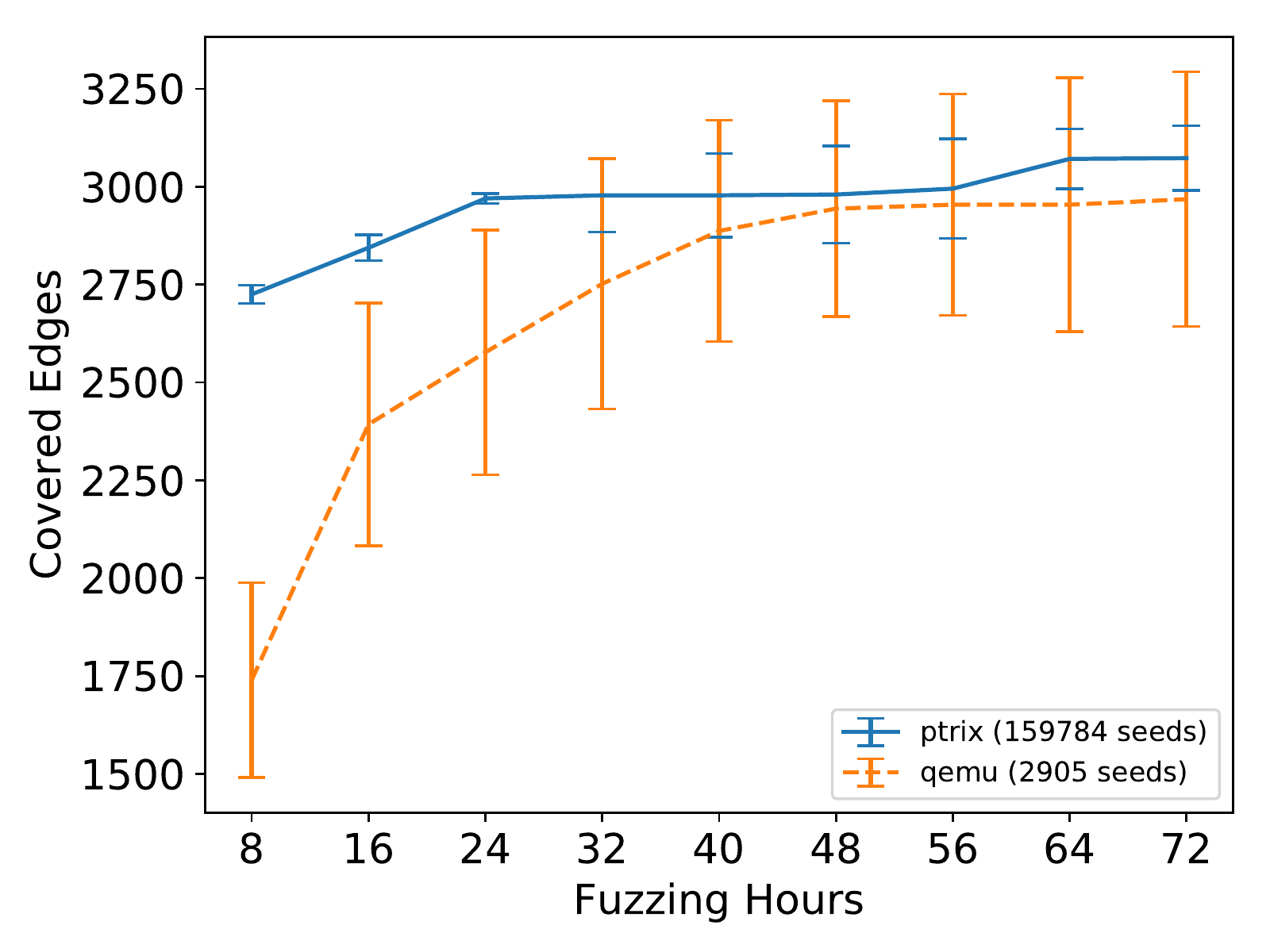}
        \caption{\scriptsize{nm}}
        \label{fig:nm}
    \end{subfigure}
    \begin{subfigure}[b]{0.325\textwidth}
        \centering
        \includegraphics[width=0.8\textwidth]{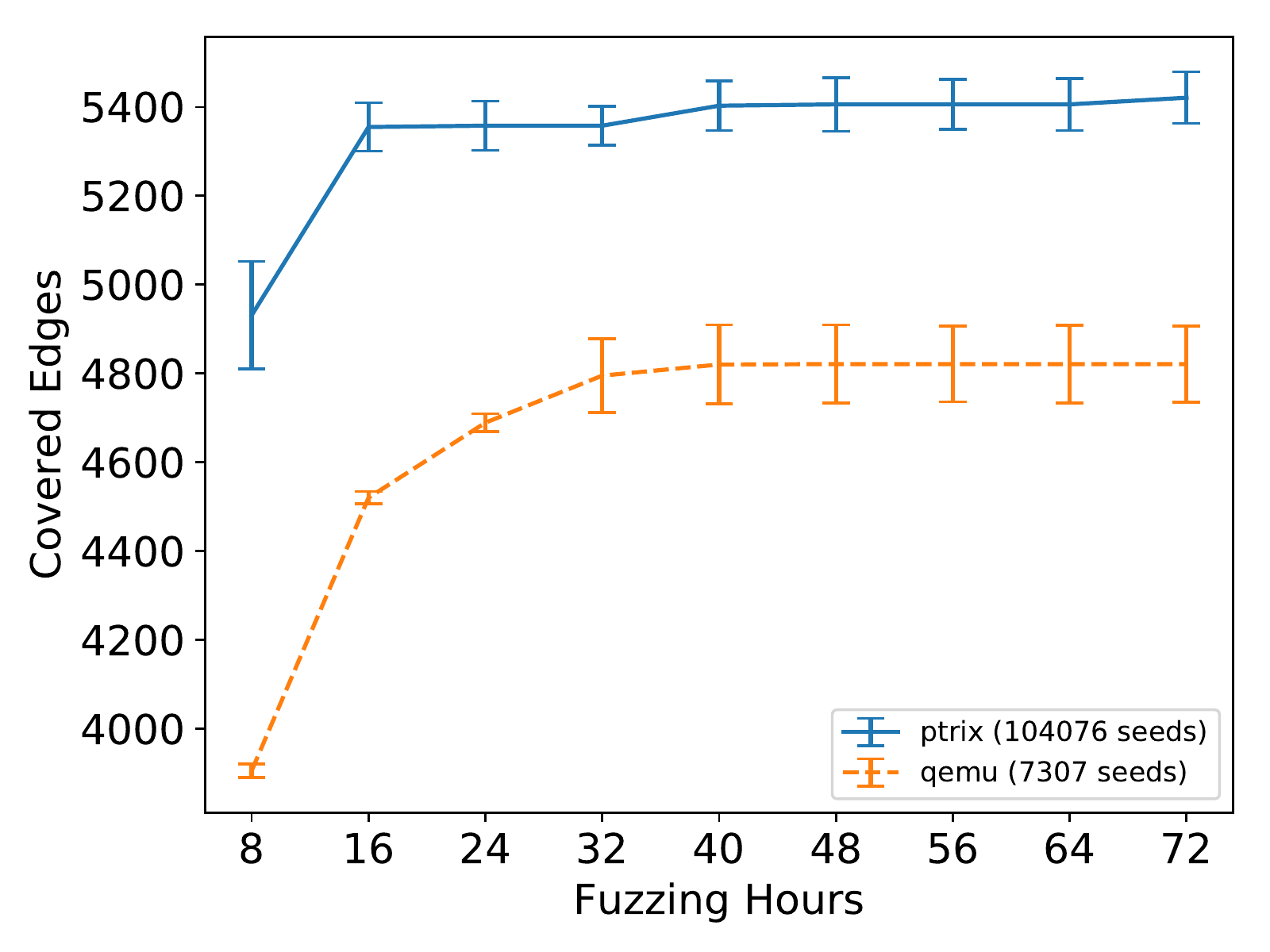}
        \caption{\scriptsize{objdump}}
        \label{fig:objdump}
    \end{subfigure}
    \caption{Edge coverage results of different fuzzing techniques for 72 hours. The star (*) besides a program name 
indicates that fuzzing on that program has saturated.}
    \label{fig:cov}
\end{figure*}

\subsection{Code Coverage Measurements}
\label{subsec:eval-cov}

As above shown, the design of \ptrix substantially accelerates the fuzzing
process. Next, we show that \ptrix is not just faster but also covers more code.
In fact, code coverage is the most widely acknowledged metric~\cite{fairfuzz, aflfast,
li2017steelix, vuzzer, skyfire, szekeres2017memory} for evaluating fuzzers. 

We run \ptrix and \afl for {\tt 72} hours or until {\tt QEMU-AFL} saturates\footnote{When
{\tt QEMU-AFL} finishes all inputs that lead to new coverage, we consider it has
saturated. The rationale is after that, {\tt QEMU-AFL} may only discover new coverage
through random attempts instead of strategic exploration.}, whichever comes
first. This long-term evaluation reduces potential random noise in results and
gives a more comprehensive view of the coverage efficiency across time. Note
that in this evaluation, we excluded {\tt Edge-pt} and {\tt PTFuzzer}. The reason is that {\tt
Edge-pt}  and {\tt PTFuzzer} explore code even slower than {\tt QEMU-AFL}, 
as echoed by our observations on the 
above {\tt 24} hour tests. 

In the following, we first present the efficiency comparison between \ptrix and
{\tt QEMU-AFL}. Then we examine the difference between code covered by the two fuzzers and
discuss the possible reasons.  

\point{Code exploration efficiency}
We calculated the code coverage using a representative quantification 
 --- \emph{number of edges between basic blocks}~\cite{fairfuzz} and summarize the results in
Figure~\ref{fig:cov}. 

As is shown in the Figure, \ptrix generally explored code space 
quicker than {\tt QEMU-AFL} across the timeline. 
Only in the case of {\tt c++filt}, \ptrix fell behind 
{\tt QEMU-AFL} from the 24th hour to the 48th hours. 
We believe this was mainly because \ptrix spent 
more time on a local code region, which is reflected 
by its increased pace after {\tt 48} hours. 
For all the {\tt 9} programs, \ptrix covered more edges than {\tt QEMU-AFL} at the end. 
In particular for {\tt objdump} and {\tt libpng}, 
\ptrix significantly increased the code coverage for over {\tt 5\%}. 
In the cases of {\tt c++filt}, {\tt nm} and {\tt mupdf}, \ptrix covered 
a similar amount of edges as {\tt QEMU-AFL} after {\tt 72} hours. 
A possible reason for \ptrix not achieving an obvious increase 
is that the fuzzers were reaching the first code coverage plateau, as their new edge discovering
rate drops almost to {\tt 0}. 

 \ptrix uses a feedback scheme with higher sensitivity, 
which tends to explore localized code more thoroughly.
By theory, this will make \ptrix move slowly around code regions. 
However, our evaluation shows an opposite conclusion.
We believe this is largely attributable to the high execution speed of \ptrix. 
This fast execution not only offsets the delay by localizing into code
regions but also accelerates the travel between different regions.
Also note that the comprehensive exploration by \ptrix is not running in vain. 
It gains new opportunities to
reach new code regions and vulnerabilities. 
We will shortly discuss this with evaluation results. 

%PS:  III.C discussion mentioned 24 hour corpus
\begin{table}[t!]
\centering
\scriptsize
\begin{tabular}{l|c|cc}
\toprule[0.5pt]
\toprule[0.5pt]

\multicolumn{1}{c|}{\bf{\emph{Program}}} & \multicolumn{3}{c}{\bf{\emph{Code coverage}}}   
 \\ \hline
 {\tt Name} & {\tt overlap} & {\tt PTRIX only} & {\tt QEMU-AFL only}
 \\ \hline

\rowcolor{mygray}

libpng & 95.60\% & 2.80\% & 1.60\%  \\

libjpeg & 89.50\% & 10.00\% & 0.50\%  \\

\rowcolor{mygray}
c++filt & 89.93\% & 5.85\% & 4.22\% \\

mupdf & 96.51\% & 2.12\% & 1.37\%  \\

\rowcolor{mygray}

{\bf Average} & 92.89\% & 5.19\% & 1.92\%   \\

\bottomrule[0.5pt]
\bottomrule[0.5pt]
\end{tabular}
\caption{Edge coverage comparison. }
\vspace{-2em}
\label{tab:case}

\vspace{-2ex}
\end{table}

\point{Code exploration effectiveness}
\ptrix and {\tt QEMU-AFL} use different feedback schemes.
Intuition suggests that the two fuzzers may explore code in
different favors. To explore this intuition, we compared the difference of edges discovered by 
\ptrix and {\tt QEMU-AFL}. Essentially, we took the union of edges from 
the two fuzzers as the baseline. Then we calculated the 
proportion that was covered by both \ptrix and {\tt QEMU-AFL}, 
by \ptrix only, and by {\tt QEMU-AFL} only. 
The average results are organized in Table~\ref{tab:case}.
We only included the cases where {\tt QEMU-AFL} has saturated. 
In those cases, {\tt QEMU-AFL} has sufficiently
expressed its exploration capability following the strategic approach, which 
enables us to better inspect whether \ptrix can really outperform {\tt QEMU-AFL}. 

As shown in the table, the two fuzzers were mostly covering the
same set of edges, but they indeed explored different code regions. 
For instance, in the case of {\tt cxxfilt}, over {\tt 10\%} of
code edges were individually discovered.
Taking a closer look, \ptrix missed significantly fewer edges than {\tt QEMU-AFL}. 
Particularly in the case of {\tt libpng}, \ptrix nearly covered 
all the edges by {\tt QEMU-AFL}. 
This indicates the path-sensitive feedback improves the code exploration of \ptrix. 
More importantly, during the long-term running, \ptrix never saturated. 
For example, when we ended the tests on {\tt c++filt}, \ptrix's pending
favorite metric was still about {\tt 1,000}. 
This demonstrated the potential of \ptrix to cover all edges that have been 
explored by {\tt QEMU-AFL}.

We have also manually inspected the different edges covered by \ptrix and {\tt QEMU-AFL}. 
Due to limited time, we have only analyzed a subset of them. 
We have identified two code regions which we believe shall only be covered by \ptrix. 
We have explained one case from {\tt libpng} in Section~\ref{sec:design} and will 
present the other case from {\tt objdump } in Section~\ref{subsec:eval-vuln}.

%15.37% of
%code edges were separately discovered by the two fuzzers with
%different feedbacks. Considering the random nature of fuzzing,
%the different code coverage might be caused by randomness.
%As a counter measurement, we added an additional evaluation.
%We re-generated the baseline by merging the code edges
%covered by two separate runs of QEMU-AFL and counted the
%above value again. The results are shown in the right part
%of Table. Those values well represent the difference due
%to randomness. Obviously randomness introduces significantly
%less difference in code coverage. This well supports that the
%new coverage scheme can guide PTRIX to discover different
%code from AFL.

\point{Code exploration comprehensiveness} 
As shown above, \ptrix and {\tt QEMU-AFL} 
may cover different code given the same amount
of time. Presumably, this is due to their different feedback schemes. 
To verify this intuition, we performed an additional analysis named
\emph{call chain analysis}. This analysis takes as inputs the corpus from \ptrix
and {\tt QEMU-AFL} in the long-term run. It re-executed each test case and collected the call
chains. A call chain is defined as follows --- When the execution reaches a leaf
node on the program's call graph, the sequence of functions on the stack is
deemed as a \emph{call chain}. The length of a call chain represents a
``locally maximal'' execution depth.

To give an overview of the call chains, we aggregated them by their lengths and
present the cumulative distribution in Figure~\ref{fig:callchain}. Generally
speaking, \ptrix produced higher proportion of shorter call chains than {\tt QEMU-AFL}. We
also observed that \ptrix usually generates the shorter call chains before the longer
ones. This shows that \ptrix spends more efforts in the beginning on shorter call
chains and then later moves onto longer ones, which is consistent with our
expectation --- \ptrix explores local code more comprehensively and does not
easily skip code paths or regions. 

\begin{table}[!ht]
\centering
\scriptsize

\begin{tabular}{l|cc|c}
\toprule[0.5pt]
\toprule[0.5pt]

\multicolumn{1}{c|}{\bf{\emph{Program}}} & \multicolumn{2}{c|}{\bf{\emph{Vulnerability Type}}}  &  \multicolumn{1}{c}{\bf{\emph{CVE}}}
 \\ \hline
	{\tt Name} & {\tt Memory Error} & {\tt DOS} &
 \\ \hline

\rowcolor{mygray}
	objdump & 4 & 1 & 0 \\

	c++filt & 3 & 2 & 2 \\

\rowcolor{mygray}
	perl & 3 & 0 & 1 \\

	nm & 4 & 1 & 0 \\ 

\rowcolor{mygray}
	gifview & 1 & 0 & 1 \\

	gdk-pixbuf & 1 & 0 & 1 \\

\rowcolor{mygray}
	nasm & 7 & 2 & 5 \\

	glibc ld & 1 & 0 & 0 \\

\rowcolor{mygray}
	libxml & 0  & 2 & 1 \\

	tcpdump & 1 & 0 & 0 \\

\rowcolor{mygray}
	unrtf & 1 & 0 & 0 \\

	libjpeg & 0 & 1 & 0 \\

\hline
%\rowcolor{mygray}
	{\bf Total} & 25 & 10 & 11 \\

\bottomrule[0.5pt]
\bottomrule[0.5pt]
\end{tabular}

\caption{Vulnerabilities discovered by \ptrix}
\label{tab:bug}
\vspace{-2.5em}
\end{table}

\subsection{Discovery of Real-world Vulnerabilities}
\label{subsec:eval-vuln}
Going beyond evaluation on fuzzing efficiency and code coverage, we further
applied \ptrix to hunt unknown bugs in the wild. We selected a set of programs
as shown in Table~\ref{tab:eval-setup} and four other well-tested programs
including {\tt gnu-ld}, {\tt curl}, {\tt nasm}, and {\tt tcpdump}. Due to
constraints of computation resources and time, we only ran each program for {\tt
24} hours. 

\ptrix triggered {\tt 19,000} unique exceptions --- unique crashes and hangs
based on the measurement of \afl. We have manually analyzed a subset of them and
confirmed \numbug new vulnerabilities. Among those vulnerabilities, {\tt
\nummembug} are memory corruptions vulnerabilities and {\tt \numdosbug}  are
Denial-of-Service (DoS) flaws that could lead to endless computation or resource
exhaustion. {\tt \numcve} CVE numbers have been created for those
vulnerabilities. We have been communicating with the developers for patches.
When those patches are available, we will disclose the details of those
vulnerabilities.

Taking a closer look at the results, we observe that the discovery of certain
vulnerabilities was indeed benefited from our new feedback. Among the 10 DoS
vulnerabilities, 9 are due to recursive calls or deep loops, which follow the
same pattern as the example shown in Figure~\ref{code:perl_bug}. As we have explained in
Section~\ref{sec:design}, {\tt QEMU-AFL} unlikely would catch them. For the memory corruption
vulnerabilities, although most of them locate in execution space that {\tt QEMU-AFL} will
also cover with high likelihood, we have identified a case that can only be
discovered using our new feedback. In the following, we the above memory corruption case and a DoS vulnerability (all the other DoS 
vulnerabilities share the same pattern). 

\textbf{Stack Overflow/Exhaustion in c++filt}. {\tt c++filt} shipped in {\tt binutils-2.29} 
can run into stack exhaustion with a long sequence of ``{\tt F}''. More
specifically, each ``{\tt F}'' leads to a recursive call chain including {\tt
demangle\_nested\_args, demangle\_args, do\_arg} and {\tt do\_type}. Stack
frames of those recursive functions gradually occupy the whole stack. 

\textbf{Integer Overflow in objdump}. In {\tt objdump} from {\tt binutils-2.29},
an integer could overflow, which further causes memory corruption. To be
specific, {\tt objdump} utilizes {\tt qsort} for sorting an array and uses the
return value of {\tt bfd\_canonicalize\_dynamic\_reloc} to specify the array
size. When exception happens, {\tt bfd\_canonicalize\_dyn\-amic\_reloc} may return
{\tt -1}. However, this is ignored by {\tt objdump} and consequently, {\tt
qsort} wrongly casts {\tt -1} to the largest unsigned value (which is taken as
the array size) and ultimately makes out-of-bound memory accesses. The {\tt
bfd\_canonicalize\_dynamic\_reloc} function implements a logic close to
Figure~\ref{code:libpng}. Because of a similar reason as we explained in Section~\ref{sec:design},
{\tt QEMU-AFL} is unable to make {\tt bfd\_canonicalize\_dynamic\_reloc} return {\tt -1}.

\section{Discussion}
\label{sec:discuss}

%drill into one long path is not necessarily bad\cite~{slowfuzz}, but consider our goal is to enlarge coverage

%high preference for quick test cases

In this section, we discuss the limitations of our current
design, insights we learned and possible future directions.

\point{Path explosion}
\ptrix implements a gray-box fuzzing scheme with path-sensitive feedback. 
This feedback metric, however, may lead to the problem of path explosion. 
That is, the fuzzer may explore a huge number of paths and correspondingly 
produce an extremely large corpus. This could further result in the exhaustion of available bitmap entries 
used by the fuzzer to record coverage. 
As we detailed in Section~\ref{sec:design}, \ptrix mitigates the path explosion problem by 
incorporating the technique of \emph{descending path sensitivity}.
This technique favors the prefix of an execution path and suppresses long paths, which   
prevents \ptrix from generating a large corpus and trapping into localized code regions. 

%\paragraph{Resource comsumption}
%We require an additional core to do parallel parsing.

\point{Generality}
\ptrix leverages PT to trace the target program. However, PT is only equipped on x86 platforms. 
We believe this will not impede the generality of the design philosophy behind \ptrix.  
Probably due to the motivation to assist debugging, hardware tracing has become a 
common feature in major architectures. Besides x86, ARM also incorporates a hardware 
feature called Embedded Trace Macrocell (ETM) to support runtime tracing. ETM, similar to PT, 
can trace the instructions with negligible performance impacts. In addition, ETM also 
provides a rich set of configuration options which can serve the requirement of \ptrix.
We, therefore, believe \ptrix can be ported to other platforms without any modifications 
to the design.

%\point{Hybrid fuzzing}
%%(edge cov plus path cov), as one is good at drilling deep path, the other is good at exploring broad path. 
%\afl supports Master-Slave mode which allows parallel execution of 
%multiple fuzzer instances and exchange of inputs between different instances. 
%This is to take advantage of the stochastic nature of fuzz-testing techniques and 
% benefit from the heterogeneity between different instances. 
%As we show in Section~\ref{sec:eval}, 
%\ptrix can explore more different --- and theoretically deeper --- code paths due to its path-sensitive feedback. 
%On the contrary, \afl is optimized to explore wider code paths. 
%Therefore, the combination of the two fuzzers is promising to cover a larger execution space 
%and we plan to work towards this direction in the future. 

\section{Related Works}
\label{sec:related}
%This research work mainly focuses on leveraging hardware to accelerate gray-box fuzzing. 
%Regarding the problem we addresses, lines of works most closely related to our own include
%gray-box fuzzing, hardware-assisted fuzzing, and improvement of fuzzing performance. 
This work focuses on leveraging PT to escalate efficiency of grey-box fuzzing on COST binaries.
With regard to this problem, the closely related research
includes binary compatible coverage-based fuzzing, improvement of coverage-based fuzzing, and combination of fuzzing and other techniques. 

\subsection{Binary Compatible Coverage-based Fuzzing}
Coverage-based fuzzing requires feedback from the target program, which can be 
obtained via lightweight program instrumentation when source code is available. 
This is, however, very challenging when only a binary is present. In the literature, 
various options have been explored. 

\subsubsection{Fuzzing with Dynamic Instrumentation} 
Dynamic instrumentation based solutions~\cite{afltech,afl-dyninst,triforce,li2017steelix} dynamically translate the binary code, the fuzzer can then intercept and collect coverage information. 
This approach, however, significantly slows down the fuzzing process. The fastest tool produced by this research line ({\tt QEMU-AFL}) reportedly introduces {\tt 2} to {\tt 5} times of overhead. 

\subsubsection{Hardware-assisted Fuzzing}
Motivated by the inefficiency of dynamic instrumentation based fuzzing systems, hardware-assisted fuzzing techniques 
were proposed recently~\cite{zhang2018ptfuzz}. Similar to \ptrix, by leveraging the newly available hardware tracing component--Intel PT~\cite{intel-pt}, {\tt Honggfuzz}~\cite{honggfuzz} and {\tt kAFL}~\cite{kafl}  
efficiently collect the execution trace from the target program. 
In contrast to \ptrix, the two systems do not fully exploit the potential of PT. {\tt Honggfuzz} only collects 
coarse-grained coverage information trading for execution throughput, which in fact degrades the code exploring capability. {\tt kAFL} and PTFuzz, however, spend too much bandwidth on reconstructing the execution flow 
from PT trace. 
 
\subsection{Improvement of Coverage-based Fuzzing}

\subsubsection{Improving Seed Generation}
%Certain works along this line focus on improving seed generation. 
Many programs take as inputs highly structured files and process these inputs over different stages ~\cite{rebert2014optimizing, driller, savior, tfuzz, bohme2017directed}. 
As a result, most randomly generated inputs will be rejected at the early stages and cannot 
reach the core logic of the target program. Therefore, based on a priori knowledge about inputs taken 
by the fuzzed programs, more targeted seeds can be generated. {\tt Skyfire}~\cite{skyfire} 
establishes a probabilistic context-sensitive grammar model by learning through a large corpus of valid inputs. 
It then uses the grammar to generate inputs that are accepted by target programs. 
Similarly, Godefroid et al. aid white-box fuzzing with a grammar-based input generator~\cite{godefroid2008grammar, 
pham2016model}. 
%Recently, Driller\cite{driller} was proposed to marry symbolic execution with fuzzing. The core idea is that symbolic execution can generate reasonable seeds that lead fuzzer through different code compartment, and fuzzer is able to fully explore each individual compartment efficiently. Currently, Driller's fuzzing component is implemented based on \afl QEMU mode, whose speed is limited by the slow dynamic instrumentation process. \ptrix, on the other hand, is a good alternative to be integrated into the Driller framework. 

\subsubsection{Improving Fuzzing Scheduling}
When there are plenty of seeds in the input queue, 
the strategy to select seeds for the following runs is very critical 
for the efficiency of fuzzing test~\cite{woo2013scheduling}. \afl~\cite{afl} develops a scheduling algorithm 
in a round-robin flavor which prefers seeds that bring new edge coverage and take less time to run. 
B{\"o}hme et al.~\cite{aflfast} propose to change that algorithm to prioritize inputs that 
follow less frequently visited paths. 
This strategy significantly accelerates the code coverage and bug discovery.
 
\subsubsection{Improving Coverage Guidance}
Providing a more informative coverage guidance is a new trend on tuning the effectiveness of
fuzz-testing techniques. CollAFL~\cite{collafl} reduced path collision introduced by \afl's 
over-approximated counted edge coverage feedback, and thus make the fuzzer more sensitive to 
new program paths. Along the same route, recent works~\cite{angora, ctxconcexec} introduced context-aware branch
coverage to decide on whether to follow inputs cover branches with new context. 
Both techniques showed that a path-based feedback is a promising 
direction to help boost fuzzer's effectiveness. \ptrix aims to provide a higher level of path
guidance, which helps \ptrix achieve high fuzzing throughput.

%\subsubsection{Taint-based fuzzing}
%Dynamic taint analysis allows reasoning of which input bytes reach
% the interesting code in the target program. 
%Many works~\cite{ganesh2009taint,bekrar2012taint,liang2013effective,haller2013dowsing,leek2007coverage} 
%are inspired by this and leverages taint analysis to improve their fuzzing techniques. 
%{\tt TaintScope}~\cite{taintscope} identifies checksum code by tainting the program inputs.
%It then makes the checksum-related data symbolic and symbolic execute the target program 
%to collect constraints along the execution path. Following that, 
%{\tt TaintScope} resolves the constraints to generate inputs that
%can pass the checksum. 
%Likewise, {\tt Vuzzer}~\cite{vuzzer} uses tainted data to discover 
%input bytes related to error-handling code and places a lower weight on such bytes when mutating inputs. 

\section{Conclusion}
\label{sec:conclude}

We present \ptrix, a binary compatible fuzz-testing tool featuring efficient code exploration capability. 
\ptrix is carefully designed and engineered to take full advantage of Intel Processor Trace as its underpinning tracing component. 
Using \ptrix, we demonstrate newly available hardware feature can significantly accelerate binary-only fuzzing through 
two elaborate designs, including a parallel scheme of trace parsing and a newly designed PT-friendly feedback.   
Also because of the new feedback provides more guidance than code coverage, 
\ptrix is able to identify \numbug new software bugs in well-tested programs that have not yet been uncovered, among them \numcve CVEs have been assigned thus far.

\section{ACKNOWLEDGMENTS}
The authors would like to thank the anonymous reviewers for their 
constructive comments. This project was supported
by the National Science Foundation 
(Grant\#: CNS-1718459, Grant\#: CNS-1748334, Grant\#: CNS-1718459)
and the Army Research Office
(Grant\#: W911NF-17-1-0039). Any opinions, findings,
and conclusions or recommendations expressed in this paper
are those of the authors and do not necessarily reflect the
views of the funding agencies.

% The next two lines define the bibliography style to be used, and the bibliography file.
{\scriptsize \bibliographystyle{ACM-Reference-Format}
\bibliography{ref}}


\begin{thebibliography}{38}

%%% the Web address.

\ifx \showCODEN    \undefined \def \showCODEN     #1{\unskip}     \fi
\ifx \showDOI      \undefined \def \showDOI       #1{#1}\fi
\ifx \showISBNx    \undefined \def \showISBNx     #1{\unskip}     \fi
\ifx \showISBNxiii \undefined \def \showISBNxiii  #1{\unskip}     \fi
\ifx \showISSN     \undefined \def \showISSN      #1{\unskip}     \fi
\ifx \showLCCN     \undefined \def \showLCCN      #1{\unskip}     \fi
\ifx \shownote     \undefined \def \shownote      #1{#1}          \fi
\ifx \showarticletitle \undefined \def \showarticletitle #1{#1}   \fi
\ifx \showURL      \undefined \def \showURL       {\relax}        \fi
% invisible to TeX
\providecommand\bibfield[2]{#2}
\providecommand\bibinfo[2]{#2}
\providecommand\natexlab[1]{#1}
\providecommand\showeprint[2][]{arXiv:#2}

\bibitem[\protect\citeauthoryear{??}{int}{2013}]%
        {intel-pt}
 \bibinfo{year}{2013}\natexlab{}.
\newblock \bibinfo{title}{Intel Processor Trace}.
\newblock
  \bibinfo{howpublished}{\url{https://software.intel.com/en-us/blogs/2013/09/18/processor-tracing}}.
\newblock


\bibitem[\protect\citeauthoryear{??}{ian}{2013}]%
        {ianareex74:online}
 \bibinfo{year}{2013}\natexlab{}.
\newblock \bibinfo{title}{Sample images for testing Exif metadata retrieval.}
\newblock \bibinfo{howpublished}{\url{https://github.com/ianare/exif-samples}}.
\newblock


\bibitem[\protect\citeauthoryear{??}{afl}{2014}]%
        {afltech}
 \bibinfo{year}{2014}\natexlab{}.
\newblock \bibinfo{title}{AFL technical details}.
\newblock
  \bibinfo{howpublished}{\url{http://lcamtuf.coredump.cx/afl/technical_details.txt}}.
\newblock


\bibitem[\protect\citeauthoryear{??}{for}{2014}]%
        {forksrv}
 \bibinfo{year}{2014}\natexlab{}.
\newblock \bibinfo{title}{Fuzzing random programs without execve()}.
\newblock
  \bibinfo{howpublished}{\url{https://lcamtuf.blogspot.com/2014/10/fuzzing-binaries-without-execve.html}}.
\newblock


\bibitem[\protect\citeauthoryear{??}{hon}{2015}]%
        {honggfuzz}
 \bibinfo{year}{2015}\natexlab{}.
\newblock \bibinfo{title}{Honggfuzz}.
\newblock \bibinfo{howpublished}{\url{http://honggfuzz.com}}.
\newblock


\bibitem[\protect\citeauthoryear{??}{afl}{2016}]%
        {afl-dyninst}
 \bibinfo{year}{2016}\natexlab{}.
\newblock \bibinfo{title}{AFL-dyninst}.
\newblock
  \bibinfo{howpublished}{\url{https://github.com/vrtadmin/moow/tree/master/afl-dyninst}}.
\newblock


\bibitem[\protect\citeauthoryear{??}{tri}{2016}]%
        {triforce}
 \bibinfo{year}{2016}\natexlab{}.
\newblock \bibinfo{title}{Project Triforce: Run AFL on Everything!}
\newblock
  \bibinfo{howpublished}{\url{https://www.nccgroup.trust/us/about-us/newsroom-and-events/blog/2016/june/project-triforce-run-afl-on-everything/}}.
\newblock


\bibitem[\protect\citeauthoryear{??}{win}{2017}]%
        {winafl}
 \bibinfo{year}{2017}\natexlab{}.
\newblock \bibinfo{title}{Harnessing Intel Processor Trace on Windows for
  fuzzing and dynamic analysis}.
\newblock
  \bibinfo{howpublished}{\url{https://recon.cx/2017/brussels/talks/intel_processor_trace.html}}.
\newblock


\bibitem[\protect\citeauthoryear{Bellard}{Bellard}{2005}]%
        {bellard2005qemu}
\bibfield{author}{\bibinfo{person}{Fabrice Bellard}.}
  \bibinfo{year}{2005}\natexlab{}.
\newblock \showarticletitle{QEMU, a Fast and Portable Dynamic Translator}. In
  \bibinfo{booktitle}{\emph{Proceedings of the Annual Conference on USENIX
  Annual Technical Conference (USENIX ATC)}}. \bibinfo{publisher}{USENIX
  Association}.
\newblock


\bibitem[\protect\citeauthoryear{B{\"o}hme, Pham, Nguyen, and
  Roychoudhury}{B{\"o}hme et~al\mbox{.}}{2017}]%
        {bohme2017directed}
\bibfield{author}{\bibinfo{person}{Marcel B{\"o}hme},
  \bibinfo{person}{Van-Thuan Pham}, \bibinfo{person}{Manh-Dung Nguyen}, {and}
  \bibinfo{person}{Abhik Roychoudhury}.} \bibinfo{year}{2017}\natexlab{}.
\newblock \showarticletitle{Directed greybox fuzzing}. In
  \bibinfo{booktitle}{\emph{Proceedings of the 2017 ACM SIGSAC Conference on
  Computer and Communications Security (CCS)}}. ACM.
\newblock


\bibitem[\protect\citeauthoryear{B{\"o}hme, Pham, and Roychoudhury}{B{\"o}hme
  et~al\mbox{.}}{2016}]%
        {aflfast}
\bibfield{author}{\bibinfo{person}{Marcel B{\"o}hme},
  \bibinfo{person}{Van-Thuan Pham}, {and} \bibinfo{person}{Abhik
  Roychoudhury}.} \bibinfo{year}{2016}\natexlab{}.
\newblock \showarticletitle{Coverage-based greybox fuzzing as markov chain}. In
  \bibinfo{booktitle}{\emph{Proceedings of the 2016 ACM SIGSAC Conference on
  Computer and Communications Security (CCS)}}. \bibinfo{publisher}{ACM}.
\newblock


\bibitem[\protect\citeauthoryear{Böck}{Böck}{2014}]%
        {TheFuzzi90:online}
\bibfield{author}{\bibinfo{person}{Hanno Böck}.}
  \bibinfo{year}{2014}\natexlab{}.
\newblock \bibinfo{title}{The Fuzzing Project - apps}.
\newblock
  \bibinfo{howpublished}{\url{https://fuzzing-project.org/software.html}}.
\newblock


\bibitem[\protect\citeauthoryear{Chen and Chen}{Chen and Chen}{2018}]%
        {angora}
\bibfield{author}{\bibinfo{person}{Peng Chen} {and} \bibinfo{person}{Hao
  Chen}.} \bibinfo{year}{2018}\natexlab{}.
\newblock \showarticletitle{Angora: Efficient Fuzzing by Principled Search}.
\newblock \bibinfo{journal}{\emph{arXiv preprint arXiv:1803.01307}}
  (\bibinfo{year}{2018}).
\newblock


\bibitem[\protect\citeauthoryear{Chen, Li, Xu, Guo, Zhou, Zhang, Wei, and
  Lu}{Chen et~al\mbox{.}}{2020}]%
        {savior}
\bibfield{author}{\bibinfo{person}{Yaohui Chen}, \bibinfo{person}{Peng Li},
  \bibinfo{person}{Jun Xu}, \bibinfo{person}{Shengjian Guo},
  \bibinfo{person}{Rundong Zhou}, \bibinfo{person}{Yulong Zhang},
  \bibinfo{person}{Tao Wei}, {and} \bibinfo{person}{Long Lu}.}
  \bibinfo{year}{2020}\natexlab{}.
\newblock \showarticletitle{SAVIOR: Towards Bug-Driven Hybrid Testing}. In
  \bibinfo{booktitle}{\emph{To appear in the 2020 IEEE Symposium on Security
  and Privacy (SP)}}. \bibinfo{publisher}{IEEE}.
\newblock


\bibitem[\protect\citeauthoryear{Gan, Zhang, Qin, Tu, Li, Pei, and Chen}{Gan
  et~al\mbox{.}}{2018}]%
        {collafl}
\bibfield{author}{\bibinfo{person}{S. Gan}, \bibinfo{person}{C. Zhang},
  \bibinfo{person}{X. Qin}, \bibinfo{person}{X. Tu}, \bibinfo{person}{K. Li},
  \bibinfo{person}{Z. Pei}, {and} \bibinfo{person}{Z. Chen}.}
  \bibinfo{year}{2018}\natexlab{}.
\newblock \showarticletitle{CollAFL: Path Sensitive Fuzzing}. In
  \bibinfo{booktitle}{\emph{2018 IEEE Symposium on Security and Privacy (SP)}}.
  IEEE.
\newblock


\bibitem[\protect\citeauthoryear{Ge, Cui, and Jaeger}{Ge et~al\mbox{.}}{2017}]%
        {griffin}
\bibfield{author}{\bibinfo{person}{Xinyang Ge}, \bibinfo{person}{Weidong Cui},
  {and} \bibinfo{person}{Trent Jaeger}.} \bibinfo{year}{2017}\natexlab{}.
\newblock \showarticletitle{Griffin: Guarding control flows using intel
  processor trace}. In \bibinfo{booktitle}{\emph{Proceedings of the
  Twenty-Second International Conference on Architectural Support for
  Programming Languages and Operating Systems (ASPLOS)}}.
  \bibinfo{publisher}{ACM}.
\newblock


\bibitem[\protect\citeauthoryear{Godefroid, Kiezun, and Levin}{Godefroid
  et~al\mbox{.}}{2008}]%
        {godefroid2008grammar}
\bibfield{author}{\bibinfo{person}{Patrice Godefroid}, \bibinfo{person}{Adam
  Kiezun}, {and} \bibinfo{person}{Michael~Y. Levin}.}
  \bibinfo{year}{2008}\natexlab{}.
\newblock \showarticletitle{Grammar-based Whitebox Fuzzing}. In
  \bibinfo{booktitle}{\emph{Proceedings of the 29th ACM SIGPLAN Conference on
  Programming Language Design and Implementation (PLDI)}}.
  \bibinfo{publisher}{ACM}.
\newblock


\bibitem[\protect\citeauthoryear{Godefroid, Levin, and Molnar}{Godefroid
  et~al\mbox{.}}{2012}]%
        {sage}
\bibfield{author}{\bibinfo{person}{Patrice Godefroid},
  \bibinfo{person}{Michael~Y. Levin}, {and} \bibinfo{person}{David Molnar}.}
  \bibinfo{year}{2012}\natexlab{}.
\newblock \showarticletitle{SAGE: Whitebox Fuzzing for Security Testing}.
\newblock \bibinfo{journal}{\emph{Queue}} \bibinfo{volume}{10},
  \bibinfo{number}{1}, Article \bibinfo{articleno}{20} (\bibinfo{date}{Jan.}
  \bibinfo{year}{2012}), \bibinfo{numpages}{8}~pages.
\newblock
\showISSN{1542-7730}


\bibitem[\protect\citeauthoryear{Henke, Schmoll, and Zseby}{Henke
  et~al\mbox{.}}{2008}]%
        {henke2008empirical}
\bibfield{author}{\bibinfo{person}{Christian Henke}, \bibinfo{person}{Carsten
  Schmoll}, {and} \bibinfo{person}{Tanja Zseby}.}
  \bibinfo{year}{2008}\natexlab{}.
\newblock \showarticletitle{Empirical Evaluation of Hash Functions for
  Multipoint Measurements}.
\newblock \bibinfo{journal}{\emph{SIGCOMM Comput. Commun. Rev.}}
  \bibinfo{volume}{38}, \bibinfo{number}{3} (\bibinfo{date}{July}
  \bibinfo{year}{2008}), \bibinfo{pages}{39--50}.
\newblock
\showISSN{0146-4833}


\bibitem[\protect\citeauthoryear{Klees, Ruef, Cooper, Wei, and Hicks}{Klees
  et~al\mbox{.}}{2018}]%
        {evalfuzz}
\bibfield{author}{\bibinfo{person}{George Klees}, \bibinfo{person}{Andrew
  Ruef}, \bibinfo{person}{Benji Cooper}, \bibinfo{person}{Shiyi Wei}, {and}
  \bibinfo{person}{Michael Hicks}.} \bibinfo{year}{2018}\natexlab{}.
\newblock \showarticletitle{Evaluating Fuzz Testing}. In
  \bibinfo{booktitle}{\emph{Proceedings of the 2018 ACM SIGSAC Conference on
  Computer and Communications Security (CCS)}}. \bibinfo{publisher}{ACM}.
\newblock


\bibitem[\protect\citeauthoryear{Labs}{Labs}{2016}]%
        {FuzzingP53:online}
\bibfield{author}{\bibinfo{person}{Geeknik Labs}.}
  \bibinfo{year}{2016}\natexlab{}.
\newblock \bibinfo{title}{Fuzzing Perl: A Tale of Two American Fuzzy Lops}.
\newblock \bibinfo{howpublished}{\url{http://www.geeknik.net/71nvhf1fp}}.
\newblock


\bibitem[\protect\citeauthoryear{lcamtuf}{lcamtuf}{2005}]%
        {afl}
\bibfield{author}{\bibinfo{person}{lcamtuf}.} \bibinfo{year}{2005}\natexlab{}.
\newblock \bibinfo{title}{american fuzzy lop}.
\newblock \bibinfo{howpublished}{\url{http://lcamtuf.coredump.cx/afl/}}.
\newblock


\bibitem[\protect\citeauthoryear{Lemieux and Sen}{Lemieux and Sen}{2017}]%
        {fairfuzz}
\bibfield{author}{\bibinfo{person}{Caroline Lemieux} {and}
  \bibinfo{person}{Koushik Sen}.} \bibinfo{year}{2017}\natexlab{}.
\newblock \showarticletitle{FairFuzz: Targeting Rare Branches to Rapidly
  Increase Greybox Fuzz Testing Coverage}.
\newblock \bibinfo{journal}{\emph{CoRR}}  \bibinfo{volume}{abs/1709.07101}
  (\bibinfo{year}{2017}).
\newblock


\bibitem[\protect\citeauthoryear{Li, Chen, Chandramohan, Lin, Liu, and Tiu}{Li
  et~al\mbox{.}}{2017}]%
        {li2017steelix}
\bibfield{author}{\bibinfo{person}{Yuekang Li}, \bibinfo{person}{Bihuan Chen},
  \bibinfo{person}{Mahinthan Chandramohan}, \bibinfo{person}{Shang-Wei Lin},
  \bibinfo{person}{Yang Liu}, {and} \bibinfo{person}{Alwen Tiu}.}
  \bibinfo{year}{2017}\natexlab{}.
\newblock \showarticletitle{Steelix: program-state based binary fuzzing}. In
  \bibinfo{booktitle}{\emph{Proceedings of the 2017 11th Joint Meeting on
  Foundations of Software Engineering (FSE)}}. ACM.
\newblock


\bibitem[\protect\citeauthoryear{Peng, Shoshitaishvili, and Payer}{Peng
  et~al\mbox{.}}{2018}]%
        {tfuzz}
\bibfield{author}{\bibinfo{person}{Hui Peng}, \bibinfo{person}{Yan
  Shoshitaishvili}, {and} \bibinfo{person}{Mathias Payer}.}
  \bibinfo{year}{2018}\natexlab{}.
\newblock \showarticletitle{T-Fuzz: fuzzing by program transformation}. In
  \bibinfo{booktitle}{\emph{2018 IEEE Symposium on Security and Privacy (SP)}}.
  IEEE.
\newblock


\bibitem[\protect\citeauthoryear{Pham, B\"{o}hme, and Roychoudhury}{Pham
  et~al\mbox{.}}{2016}]%
        {pham2016model}
\bibfield{author}{\bibinfo{person}{Van-Thuan Pham}, \bibinfo{person}{Marcel
  B\"{o}hme}, {and} \bibinfo{person}{Abhik Roychoudhury}.}
  \bibinfo{year}{2016}\natexlab{}.
\newblock \showarticletitle{Model-based Whitebox Fuzzing for Program Binaries}.
  In \bibinfo{booktitle}{\emph{Proceedings of the 31st IEEE/ACM International
  Conference on Automated Software Engineering (ASE)}}.
  \bibinfo{publisher}{ACM}.
\newblock


\bibitem[\protect\citeauthoryear{Rawat, Jain, Kumar, Cojocar, Giuffrida, and
  Bos}{Rawat et~al\mbox{.}}{2017}]%
        {vuzzer}
\bibfield{author}{\bibinfo{person}{Sanjay Rawat}, \bibinfo{person}{Vivek Jain},
  \bibinfo{person}{Ashish Kumar}, \bibinfo{person}{Lucian Cojocar},
  \bibinfo{person}{Cristiano Giuffrida}, {and} \bibinfo{person}{Herbert Bos}.}
  \bibinfo{year}{2017}\natexlab{}.
\newblock \showarticletitle{Vuzzer: Application-aware evolutionary fuzzing}. In
  \bibinfo{booktitle}{\emph{Proceedings of the Network and Distributed System
  Security Symposium (NDSS)}}.
\newblock


\bibitem[\protect\citeauthoryear{Rebert, Cha, Avgerinos, Foote, Warren, Grieco,
  and Brumley}{Rebert et~al\mbox{.}}{2014}]%
        {rebert2014optimizing}
\bibfield{author}{\bibinfo{person}{Alexandre Rebert}, \bibinfo{person}{Sang~Kil
  Cha}, \bibinfo{person}{Thanassis Avgerinos}, \bibinfo{person}{Jonathan
  Foote}, \bibinfo{person}{David Warren}, \bibinfo{person}{Gustavo Grieco},
  {and} \bibinfo{person}{David Brumley}.} \bibinfo{year}{2014}\natexlab{}.
\newblock \showarticletitle{Optimizing Seed Selection for Fuzzing}. In
  \bibinfo{booktitle}{\emph{Proceedings of the 23rd USENIX Conference on
  Security Symposium (USENIX Security)}}. \bibinfo{publisher}{USENIX
  Association}.
\newblock


\bibitem[\protect\citeauthoryear{Schumilo, Aschermann, Gawlik, Schinzel, and
  Holz}{Schumilo et~al\mbox{.}}{2017}]%
        {kafl}
\bibfield{author}{\bibinfo{person}{Sergej Schumilo}, \bibinfo{person}{Cornelius
  Aschermann}, \bibinfo{person}{Robert Gawlik}, \bibinfo{person}{Sebastian
  Schinzel}, {and} \bibinfo{person}{Thorsten Holz}.}
  \bibinfo{year}{2017}\natexlab{}.
\newblock \showarticletitle{kAFL: Hardware-Assisted Feedback Fuzzing for {OS}
  Kernels}. In \bibinfo{booktitle}{\emph{Proceedings of the 26rd USENIX
  Conference on Security Symposium (USENIX Security)}}.
  \bibinfo{publisher}{{USENIX} Association}.
\newblock


\bibitem[\protect\citeauthoryear{Seltzer and Yigit}{Seltzer and Yigit}{1991}]%
        {seltzer1991new}
\bibfield{author}{\bibinfo{person}{Margo~I Seltzer} {and} \bibinfo{person}{Ozan
  Yigit}.} \bibinfo{year}{1991}\natexlab{}.
\newblock \showarticletitle{A New Hashing Package for UNIX.}. In
  \bibinfo{booktitle}{\emph{USENIX Winter}}. \bibinfo{publisher}{USENIX}.
\newblock


\bibitem[\protect\citeauthoryear{Seo and Kim}{Seo and Kim}{2014}]%
        {ctxconcexec}
\bibfield{author}{\bibinfo{person}{Hyunmin Seo} {and} \bibinfo{person}{Sunghun
  Kim}.} \bibinfo{year}{2014}\natexlab{}.
\newblock \showarticletitle{How we get there: A context-guided search strategy
  in concolic testing}. In \bibinfo{booktitle}{\emph{Proceedings of the 22nd
  ACM SIGSOFT International Symposium on Foundations of Software Engineering
  (FSE)}}. ACM.
\newblock


\bibitem[\protect\citeauthoryear{Shoshitaishvili, Wang, Salls, Stephens,
  Polino, Dutcher, Grosen, Feng, Hauser, Kruegel, and Vigna}{Shoshitaishvili
  et~al\mbox{.}}{2016}]%
        {angr}
\bibfield{author}{\bibinfo{person}{Yan Shoshitaishvili}, \bibinfo{person}{Ruoyu
  Wang}, \bibinfo{person}{Christopher Salls}, \bibinfo{person}{Nick Stephens},
  \bibinfo{person}{Mario Polino}, \bibinfo{person}{Audrey Dutcher},
  \bibinfo{person}{John Grosen}, \bibinfo{person}{Siji Feng},
  \bibinfo{person}{Christophe Hauser}, \bibinfo{person}{Christopher Kruegel},
  {and} \bibinfo{person}{Giovanni Vigna}.} \bibinfo{year}{2016}\natexlab{}.
\newblock \showarticletitle{{SoK: (State of) The Art of War: Offensive
  Techniques in Binary Analysis}}. In \bibinfo{booktitle}{\emph{Proceedings of
  the 2016 IEEE Symposium on Security and Privacy (SP)}}.
  \bibinfo{publisher}{IEEE}.
\newblock


\bibitem[\protect\citeauthoryear{Stephens, Grosen, Salls, Dutcher, Wang,
  Corbetta, Shoshitaishvili, Kruegel, and Vigna}{Stephens
  et~al\mbox{.}}{2016}]%
        {driller}
\bibfield{author}{\bibinfo{person}{Nick Stephens}, \bibinfo{person}{John
  Grosen}, \bibinfo{person}{Christopher Salls}, \bibinfo{person}{Andrew
  Dutcher}, \bibinfo{person}{Ruoyu Wang}, \bibinfo{person}{Jacopo Corbetta},
  \bibinfo{person}{Yan Shoshitaishvili}, \bibinfo{person}{Christopher Kruegel},
  {and} \bibinfo{person}{Giovanni Vigna}.} \bibinfo{year}{2016}\natexlab{}.
\newblock \showarticletitle{Driller: Augmenting Fuzzing Through Selective
  Symbolic Execution.}. In \bibinfo{booktitle}{\emph{Proceedings of the Network
  and Distributed System Security Symposium (NDSS)}}.
\newblock


\bibitem[\protect\citeauthoryear{Szekeres}{Szekeres}{2017}]%
        {szekeres2017memory}
\bibfield{author}{\bibinfo{person}{L{\'a}szl{\'o} Szekeres}.}
  \bibinfo{year}{2017}\natexlab{}.
\newblock \emph{\bibinfo{title}{Memory corruption mitigation via hardening and
  testing}}.
\newblock \bibinfo{thesistype}{Ph.D. Dissertation}. \bibinfo{school}{Stony
  Brook University}.
\newblock


\bibitem[\protect\citeauthoryear{Wang, Chen, Wei, and Liu}{Wang
  et~al\mbox{.}}{2017}]%
        {skyfire}
\bibfield{author}{\bibinfo{person}{Junjie Wang}, \bibinfo{person}{Bihuan Chen},
  \bibinfo{person}{Lei Wei}, {and} \bibinfo{person}{Yang Liu}.}
  \bibinfo{year}{2017}\natexlab{}.
\newblock \showarticletitle{Skyfire: Data-driven seed generation for fuzzing}.
  In \bibinfo{booktitle}{\emph{Proceedings of the 2017 IEEE Symposium on
  Security and Privacy (SP)}}. \bibinfo{publisher}{IEEE}.
\newblock


\bibitem[\protect\citeauthoryear{Woo, Cha, Gottlieb, and Brumley}{Woo
  et~al\mbox{.}}{2013}]%
        {woo2013scheduling}
\bibfield{author}{\bibinfo{person}{Maverick Woo}, \bibinfo{person}{Sang~Kil
  Cha}, \bibinfo{person}{Samantha Gottlieb}, {and} \bibinfo{person}{David
  Brumley}.} \bibinfo{year}{2013}\natexlab{}.
\newblock \showarticletitle{Scheduling black-box mutational fuzzing}. In
  \bibinfo{booktitle}{\emph{Proceedings of the 2013 ACM SIGSAC conference on
  Computer and Communications Security (CCS)}}. ACM.
\newblock


\bibitem[\protect\citeauthoryear{Xu, Kashyap, Min, and Kim}{Xu
  et~al\mbox{.}}{2017}]%
        {designing}
\bibfield{author}{\bibinfo{person}{Wen Xu}, \bibinfo{person}{Sanidhya Kashyap},
  \bibinfo{person}{Changwoo Min}, {and} \bibinfo{person}{Taesoo Kim}.}
  \bibinfo{year}{2017}\natexlab{}.
\newblock \showarticletitle{Designing New Operating Primitives to Improve
  Fuzzing Performance}. In \bibinfo{booktitle}{\emph{Proceedings of the 2017
  ACM SIGSAC Conference on Computer and Communications Security (CCS)}}.
  \bibinfo{publisher}{ACM}.
\newblock


\bibitem[\protect\citeauthoryear{Zhang, Zhou, Luo, Wu, and Min}{Zhang
  et~al\mbox{.}}{2018}]%
        {zhang2018ptfuzz}
\bibfield{author}{\bibinfo{person}{Gen Zhang}, \bibinfo{person}{Xu Zhou},
  \bibinfo{person}{Yingqi Luo}, \bibinfo{person}{Xugang Wu}, {and}
  \bibinfo{person}{Erxue Min}.} \bibinfo{year}{2018}\natexlab{}.
\newblock \showarticletitle{PTfuzz: Guided Fuzzing with Processor Trace
  Feedback}.
\newblock \bibinfo{journal}{\emph{IEEE Access}} (\bibinfo{year}{2018}).
\newblock


\end{thebibliography}
\newpage
\begin{appendices}
\section{Supplementary Figures and Evaluation Data}
\label{sec:appendix1}

\begin{figure}[ht]
\includegraphics[scale=0.49]{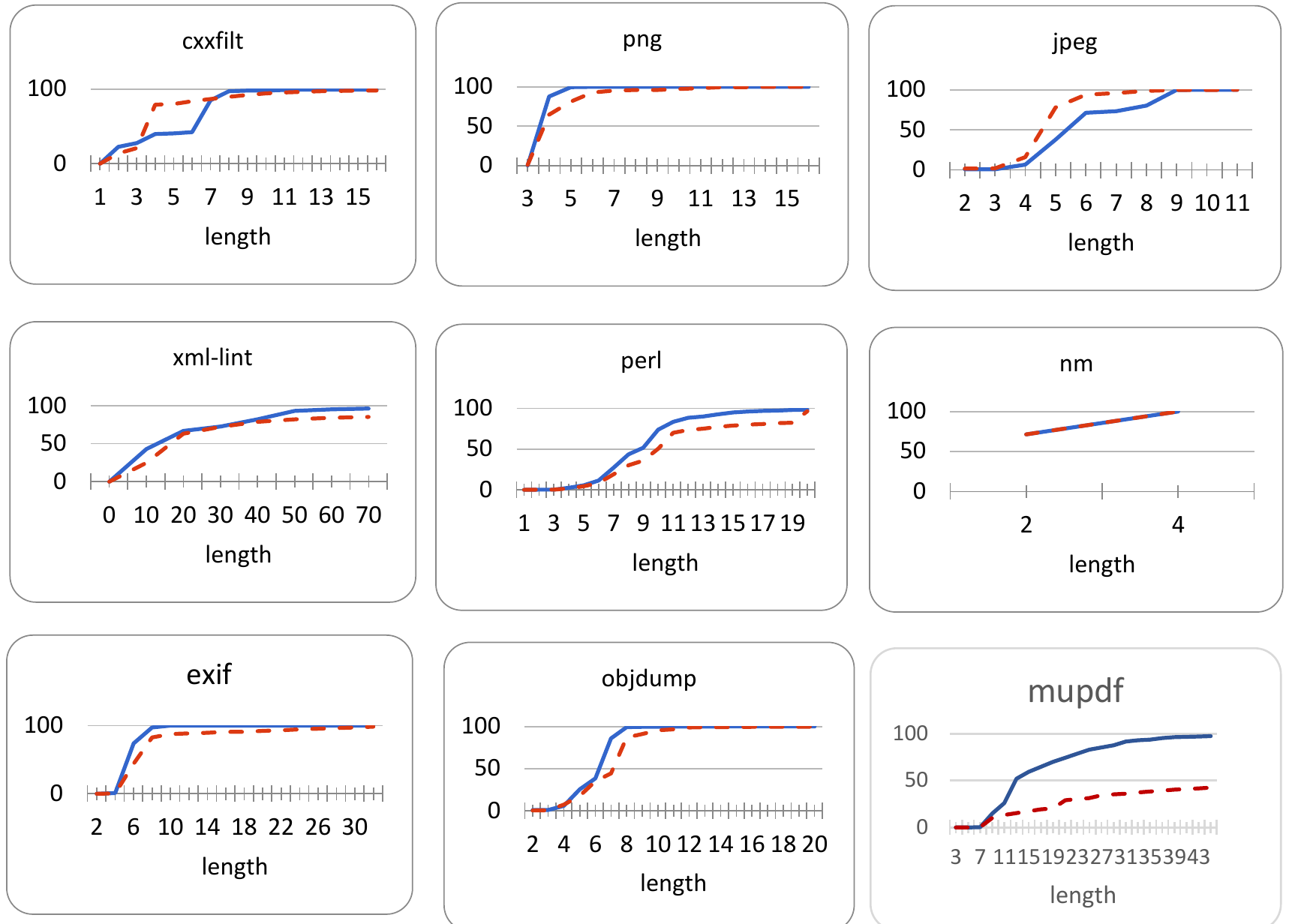}
\caption{CDF of Call chains triggered by different fuzzing techniques. PTrix (Solidline), QEMU-AFL (Dashline)}
\label{fig:callchain}
\end{figure}

\end{appendices}
% 
% If your work has an appendix, this is the place to put it.
%\appendix

\end{document}